\documentclass[twocolumn,pra,amsmath,amssymb,superscriptaddress,longbibliography,nofootinbib]{revtex4-2}

\usepackage{lipsum}
\usepackage[T1]{fontenc}
\usepackage{bbold}

\usepackage{dsfont}

\usepackage{color} 
\usepackage{ulem}
\usepackage{physics, calc} 
\usepackage{graphicx}
\usepackage{enumitem}
\usepackage[ruled,vlined]{algorithm2e}
\usepackage[table]{xcolor}
\usepackage{array}
\usepackage[citecolor=blue, urlcolor=black, linkcolor=blue, colorlinks=true]{hyperref}

\usepackage{tikz}
\usetikzlibrary{external,decorations.markings}

\makeatletter
\newcommand*\rel@kern[1]{\kern#1\dimexpr\macc@kerna}
\newcommand*\widebar[1]{%
  \begingroup
  \def\mathaccent##1##2{%
    \rel@kern{0.8}%
    \overline{\rel@kern{-0.8}\macc@nucleus\rel@kern{0.2}}%
    \rel@kern{-0.2}%
  }%
  \macc@depth\@ne
  \let\math@bgroup\@empty \let\math@egroup\macc@set@skewchar
  \mathsurround\z@ \frozen@everymath{\mathgroup\macc@group\relax}%
  \macc@set@skewchar\relax
  \let\mathaccentV\macc@nested@a
  \macc@nested@a\relax111{#1}%
  \endgroup
}
\makeatother

\newtheorem{result}{Result}

\usepackage{framed}
\definecolor{shadecolor}{gray}{0.95}
\usepackage{mathtools}

\usepackage[ruled,vlined]{algorithm2e}
\usepackage{subfigure}

\makeatletter
\def\l@subsection#1#2{}
\def\l@subsubsection#1#2{}
\makeatother

\begin{document}

\author{Nick G. Jones}
\affiliation{Mathematical Institute, University of Oxford, Oxford, OX2 6GG, UK}
\affiliation{The Heilbronn Institute for Mathematical Research, Bristol, UK}
\author{Julian Bibo}
\affiliation{Department of Physics, TFK, Technische Universit{\"a}t M{\"u}nchen, James-Franck-Stra{\ss}e 1, D-85748 Garching, Germany}
\affiliation{Munich Center for Quantum Science and Technology (MQCST), D-80799 Munich, Germany}
\author{Bernhard Jobst}
\affiliation{Department of Physics, TFK, Technische Universit{\"a}t M{\"u}nchen, James-Franck-Stra{\ss}e 1, D-85748 Garching, Germany}
\author{Frank Pollmann}
\affiliation{Department of Physics, TFK, Technische Universit{\"a}t M{\"u}nchen, James-Franck-Stra{\ss}e 1, D-85748 Garching, Germany}
\affiliation{Munich Center for Quantum Science and Technology (MQCST), D-80799 Munich, Germany}
\author{Adam Smith}
\affiliation{Department of Physics, TFK, Technische Universit{\"a}t M{\"u}nchen, James-Franck-Stra{\ss}e 1, D-85748 Garching, Germany}
\affiliation{School of Physics and Astronomy, University of Nottingham, Nottingham, NG7 2RD, UK}
\affiliation{Centre for the Mathematics and Theoretical Physics of Quantum Non-Equilibrium Systems, University of Nottingham, Nottingham, NG7 2RD, UK}
\author{Ruben Verresen}
\affiliation{Department of Physics, Harvard University, Cambridge, MA 02138, USA}

\title{Skeleton of Matrix-Product-State-Solvable Models\texorpdfstring{\\}{ }Connecting Topological Phases of Matter}

\normalem

\begin{abstract}
Models whose ground states can be written as an exact matrix product state (MPS) provide valuable insights into phases of matter.
While MPS-solvable models are typically studied as isolated points in a phase diagram, they can belong to a connected network of MPS-solvable models, which we call the \emph{MPS skeleton}. As a case study where we can completely unearth this skeleton, we focus on the one-dimensional BDI class---non-interacting spinless fermions with time-reversal symmetry. This class, labelled by a topological winding number, contains the Kitaev chain and is Jordan-Wigner-dual to various symmetry-breaking and symmetry-protected topological (SPT) spin chains.
We show that one can read off from the Hamiltonian whether its ground state is an MPS:
defining a polynomial whose coefficients are the Hamiltonian parameters, MPS-solvability corresponds to this polynomial being a perfect square.
We provide an explicit construction of the ground state MPS, its bond dimension growing exponentially with the range of the Hamiltonian. This complete characterization of the MPS skeleton in parameter space has three significant consequences: (i) any two topologically distinct phases in this class admit a path of MPS-solvable models between them, including the phase transition which obeys an area law for its entanglement entropy; (ii) we illustrate that the subset of MPS-solvable models is dense in this class by constructing a sequence of MPS-solvable models which converge to the Kitaev chain (equivalently, the quantum Ising chain in a transverse field); (iii) a subset of these MPS states can be particularly efficiently processed on a noisy intermediate-scale quantum computer.
\end{abstract}

\maketitle

\tableofcontents

\normalem 
\vspace{0.2cm}
\footnotesize{The published version of this article is Phys. Rev. Research 3, 033265; \href{https://doi.org/10.1103/PhysRevResearch.3.033265}{https://doi.org/10.1103/PhysRevResearch.3.033265}.}
\normalsize
\section{Introduction \label{sec:intro}}
The realization that the entanglement of gapped many-body ground states obeys an area law was a breakthrough for condensed matter physics~\cite{Hastings2007}. It justifies the use of tensor network states as a description of the wavefunction, having become a key analytic and numerical tool \cite{White1992a,Fannes89,Schollwoeck11,Stoudenmire12,Hauschild18,Paeckel19,Zeng19,Cirac2020}. These tools are most refined for the case of matrix product states (MPS) describing one-dimensional systems. In most scenarios, such MPS are \emph{approximations} to the true ground states. However, a wide variety of Hamiltonians are known where the ground state is an exact MPS---i.e., with a finite bond dimension in the thermodynamic limit \cite{Majumdar69,Affleck1987,Scalapino98,Raussendorf01,Wolf06,Greiter07,Schuricht08,Tu08,Alet11,Nonne13,Morimoto14,Geraedts14,Roy18,Gozel19}. 

The importance of such models is well-illustrated by the discovery of the Affleck-Kennedy-Lieb-Tasaki (AKLT) spin-$1$ chain in 1987~\cite{Affleck1987}. This model (itself inspired by the Majumdar-Ghosh spin-$1/2$ chain \cite{Majumdar69} and Haldane's conjecture \cite{Haldane83a,Haldane83b}) led to the development and discovery of both MPS \cite{Fannes89,Cirac2020} and symmetry-protected topological (SPT) phases of matter \cite{Affleck88,Kennedy90,Kennedy92,Ng94,Berg08,Gu09,Pollmann10,Turner11,Chen11,Schuch11,Pollmann12}. In particular, through the study of its MPS, it was realized that its degenerate edge modes and entanglement levels are protected by, e.g., spin-rotation or time-reversal symmetry \cite{Berg08,Gu09,Pollmann10}. In fact, any one-dimensional (1D) SPT phase admits a fixed-point MPS state \cite{Chen11,Kapustin14}. Despite its importance, the AKLT model is commonly thought of as an \emph{isolated} MPS-solvable point in parameter space.

Less explored are continuous \emph{paths} of MPS-solvable models connecting distinct phases of matter through a quantum phase transition.
One option is to simply define paths in the manifold of exact MPS states in Hilbert space. Indeed, using the well-established parent Hamiltonian construction, this gives a path of MPS-solvable models \cite{Schuch11,Chen11,Cirac2020}. While MPS cannot capture conformal critical points which have diverging entanglement entropy, they can describe certain multicritical points where the gap closes \cite{Wolf06}. Paths approaching such points can exhibit a diverging correlation length in a finite bond dimension MPS. Explicit discussions in the literature of such instances seem to be rare, an example being the disorder line 
in the spin-$1/2$ XY chain \cite{Barouch71,Kurmann82,Taylor83,Mueller85,Chung01,Franchini07}; this interpolates between two distinct ferromagnets by passing through a multicritical point\footnote{We note that this path is unusual since the two symmetry-broken ground states are a product state throughout: $\otimes_n \left( \cos \theta | \uparrow\rangle_n \pm \sin \theta |\downarrow \rangle_n \right)$ (see, for example, Ref.~\cite{Franchini07}). This becomes a unique symmetry-preserving product state at the multicritical point as $\theta \to 0$. In other examples, the correlation length can diverge.} with dynamic critical exponent $z=2$.
A reincarnation of this example---related by a Kramers-Wannier transformation---is the MPS path connecting the trivial phase to the Haldane phase as realized by the cluster state \cite{Wolf06,Smith19}. 
Let us note that the aforementioned parent Hamiltonian construction is not unique and can give rise to unwieldy Hamiltonians which are not necessarily in a class of interest.

In this work, we do not start from a path of MPS: instead, we specify the Hamiltonian class and ask which models have an MPS ground state. This leads to MPS-solvable paths forming the skeleton around which the rest of the phase diagram is structured. For a particular class of non-interacting symmetric Hamiltonians (BDI class \cite{Altland97}), we develop a general understanding of paths of MPS-solvable models, connecting the distinct SPT phases. Different paths connect at joints where the system is multicritical and still has an MPS ground state. We refer to this network as \textit{the MPS skeleton}.
Remarkably, this skeleton is dense in this class (similar to how rational numbers are dense on the real line): any gapped ground state can be obtained as a sequence of Hamiltonians whose ground state is an exact MPS.

We note that while the idea of the MPS skeleton is by no means particular to non-interacting systems, this setting is an interesting case study. Despite free-fermion Hamiltonians and MPS-solvable systems both being pinnacles of solubility, they have a rich interplay: one cannot typically write the ground state of a free-fermion system as an exact MPS due to its entanglement spectrum having infinite rank\footnote{For an explicit example, the entanglement spectrum is calculated analytically for the XY-model in \cite{Franchini2010}.}, and there is no analytic handle on truncating this to a particular bond dimension. This truncation has been investigated numerically: an approach for the XY model is given in Ref.~\cite{Rams_2015}; and more generally there are approaches based on truncating the free-fermion correlation matrix \cite{Fishman15,Schuch19}, the `MPO-MPS method' \cite{Jin20} and through Schmidt decomposition \cite{Petrica21}.
The MPS description of free-fermion states has been explored before in the context of Gaussian MPS \cite{Schuch08,Kraus10,Schuch19}. Using this framework, Ref.~\cite{Kraus10} showed that free-fermion states admitting an exact (Gaussian) MPS representation have a correlation matrix that satisfies a certain property \cite{Cirac2020}, readily applying to arbitrary dimensions. 
We will see that, for the BDI class, this property coincides with our characterization of the MPS skeleton. Indeed, our analysis shows that the implication also works the other way: this property is sufficient for MPS-solvability, and moreover we give an explicit construction of the ground state. We do not appeal to the formalism of Gaussian MPS, and it would be interesting to translate our results into that language, giving a concrete subclass of Gaussian MPS for which we have an explicit construction.

To briefly outline the paper, in Section~\ref{sec:results} we introduce BDI Hamiltonians and summarize our main results, followed by explicit examples in Section~\ref{sec:example}. We provide the derivation of our results in Section~\ref{sec:analysis} and elaborate on special cases in Section~\ref{sec:special}.

\section{Summary of main results \label{sec:results}}
\subsection{Model}\label{sec:model}
We consider a chain of Majorana operators $\gamma_n$ and $\tilde \gamma_n$, which are respectively real and imaginary under time-reversal\footnote{More precisely, if $c_n$ is a complex fermionic operator, the Majorana operators are $\gamma_n = c_n + c_n^\dagger$ and $\tilde \gamma_n = i\left( c_n^\dagger-c_n \right)$. Then $\{ \gamma_n, \tilde \gamma_m \} = 0$ and $\{ \gamma_n, \gamma_m \} = 2 \delta_{nm}$.}.
The most general Hamiltonian term in the free-fermion BDI class is $h_{n,\alpha} = i \tilde \gamma_n \gamma_{n+\alpha}$, which has the convenient property $ h_{n,\alpha}^2=1$. Any translation-invariant BDI Hamiltonian can be written as
\begin{align}
    H = \frac{1}{2} \sum_{n,\alpha} t_\alpha h_{n,\alpha} \label{eq:H}
\end{align}
where $t_\alpha \in \mathbb R$ due to hermiticity. The special cases $H_\omega$ where $t_\alpha = \delta_{\alpha,\omega}$ are stabilizer code Hamiltonians (all terms commute), and are fixed-point Hamiltonians in the phase with winding number $\omega \in \mathbb Z$. Note that $\omega=1$ gives the Kitaev chain \cite{Kitaev2001}.

It is convenient to encode the information of the Hamiltonian parameters in a Laurent\footnote{Note that it can contain negative powers of $z$ if $t_\alpha \neq0$ for some negative $\alpha$.} polynomial
\begin{align}
    f(z) = \sum_\alpha t_\alpha z^\alpha. \label{eq:fz}
\end{align}
Previous work has already shown that a multitude of physical information can be readily extracted from $f(z)$. 
E.g., the single-particle spectrum is given by $\varepsilon_k = |f(e^{ik})|$; the correlation length is $\xi = \max_i \{ 1/|\ln|\zeta_i|| \}$ where $\zeta_i$ are the roots of $f(z)$; the topological invariant $\omega=N_z-N_p$, where $N_z$ is the number of roots inside the unit circle and $N_p$ is the degree of the pole at $z=0$ \cite{Verresen18,Jones2019}. We will refer to the ground state of $f(z)$: this means the ground state of the related Hamiltonian.

Under the usual Jordan-Wigner transformation (see Eq.~\eqref{eq:JW}), the Hamiltonian terms become
\begin{equation}
    h_{n,\alpha} = \left\{
    \begin{array}{ccc}
    - X_n Z_{n+1} \cdots Z_{n+\alpha-1} X_{n+\alpha} & & \textrm{if }\alpha>0 \\
    Z_n & & \textrm{if }\alpha = 0\\
    - Y_{n+\alpha} Z_{n+\alpha+1} \cdots Z_{n-1} Y_{n} & & \textrm{if }\alpha<0
    \end{array}. \right.\label{eq:hnspin}
\end{equation}
In particular, the fixed-point Hamiltonians $H_\omega$ correspond to symmetry-breaking or SPT\footnote{E.g., $H_1$ has symmetry-breaking order, $H_2$ has SPT order, and $H_3$ has \emph{both}; see Ref.~\cite{Verresen17} for details.} Hamiltonians such as the Ising model $H_1 = -\frac{1}{2} \sum_n X_n X_{n+1}$ and cluster model $H_2 = -\frac{1}{2} \sum_n X_{n-1} Z_n X_{n+1}$. More generally, $H_\alpha$ is a generalized cluster model, or $\alpha$-chain~\cite{Suzuki71,Keating2004,Smacchia11,Ohta16,Verresen17}.

\subsection{The MPS skeleton}
Here, we add to this body of knowledge by characterizing when the ground state of Eq.~\eqref{eq:H} is an MPS. 
For our purposes, this means that we have an explicit finite-depth circuit representation for the ground state (note that the gates of this circuit will not necessarily be unitary). In Section \ref{sec:MPS} we will make the connection to the usual definition of an MPS \cite{Cirac2020} as a tensor network where the ground state is a contraction of tensors with virtual indices that have bond dimension $\chi$. We have that the ground state is an exact MPS if $f(z)$ is a square; more precisely:
\begin{shaded}
\begin{result}[Existence of MPS.]\label{result1}
If $f(z) =  z^p g(z)^2$ for $p \in \mathbb Z$ and $g(z) = \sum_{k= 0}^d s_k z^k$ (with $s_k \in \mathbb R$), then the Hamiltonian given in Eq.~\eqref{eq:H} is frustration-free. Moreover, its ground state can be exactly represented as an MPS with finite bond dimension $\chi$. If we ensure that $s_0 \neq 0 \neq s_d$ (which one can always do by appropriately choosing $p$ and $d \geq 0$), 
then
\begin{align}
    \log_2 \chi = \lceil\mathrm{range}(H)/2\rceil \label{eq:chi}
\end{align}
where $\mathrm{range}(H)$ is defined as the largest power of either $z$ or $1/z$ in $f(z)$ 
and $\lceil \cdot \rceil : \mathbb{R} \rightarrow \mathbb{Z}$ is the ceiling function. 
\end{result}
\end{shaded}
The formula for $\chi$ is for MPS which are symmetric under fermion parity (or, equivalently, spin-flip symmetry); in the case of spontaneous symmetry breaking (for the spin chain), this formula applies to the cat state, whereas the symmetry-broken state has $ \log_2 \chi = \lfloor\textrm{range}(H)/2\rfloor$. We believe that this formula for $\chi$ is generically optimal, and in certain cases this can be proved---this is discussed in Section \ref{sec:MPS}. 

This gives a complete characterization of all BDI Hamiltonians with an exact MPS ground state (with finite $\chi$ in the thermodynamic limit). More precisely, the above result holds for the broader class of Hamiltonians $f(z) = \pm z^p g(z)^2 h(z)$, where $h(z)$ is any Laurent polynomial that satisfies $h(1/z) = h(z)$, has no roots on the unit circle and has positive constant term. However, (i) the sign can easily be toggled (see Eq.~\eqref{eq:sigmatoggle}) and so we will henceforth consider a positive sign, and (ii) the ground state is independent of $h(z)$. Any $f(z)$ \emph{not} of this form has correlation functions with asymptotics containing terms of the form $N^{\alpha}\exp(-N/\xi)$ for $\alpha \notin \mathbb{N}_{+}$ and $\xi \leq \infty$, which cannot be captured by an MPS with a finite bond dimension\footnote{The case $\xi=\infty$ corresponds to a critical system.}. These claims are proved in Appendix~\ref{app:BDI}. 

We point out that a similar necessary condition for MPS-solvability was obtained in Ref.~\cite{Schuch08} in the context of lattice models for free (bosonic) oscillators; the argument straightforwardly extends to the fermionic case\footnote{We are grateful to N. Schuch for clarifying this point.}. A relation was also given between the bond dimension and the interaction range of the Hamiltonian. The fermionic analogue of these Gaussian MPS are discussed in Refs.~\cite{Kraus10,Cirac2020}; a characteristic of such states is that rational functions of $z=e^{ik}$ generate correlations by Fourier transform. Indeed, for the case under discussion, correlations are generated by $\sqrt{f(z)/f(1/z)}$ which, on the MPS skeleton, reduces to the rational function $z^pg(z)/g(1/z)$. Our results thus show that this characterization is sufficient as well as necessary. Moreover, our proof is constructive---we will turn to this now.

\subsection{Construction of MPS}
Excluding a measure zero set, we have a closed form for the MPS wavefunction. This is most easily described in terms of $d$ real parameters $b_{k=1,\cdots,d}$ that are obtained from the following recursion. 
Here $d$ is the degree of $g(z)$ and $(s_0,\cdots,s_d)$ are its coefficients, as defined in Result \ref{result1}; writing $\vec s = (s_0,s_1,\cdots,s_d)$ and $\textrm{flip}(\vec s) = (s_d,s_{d-1},\cdots,s_0)$ we have:
\begin{algorithm}[H]
 \label{eq:recursion}
 \For{$k=d,\cdots,2,1$}{
  $b_k = s_k/s_0$\;
  $\vec s = \vec s - b_k \times \textrm{flip}(\vec s)$\;
  drop last entry of $\vec s$\;
 }
 \caption{obtaining $b_k$ from $\vec s$}
\end{algorithm}
The outcome of this algorithm will only be used if $|b_k| \neq 1$ for all $k$ (here we thus exclude a measure zero case). Given this condition, one can show that $s_0 \neq 0$ at each step, ensuring that the ratio $s_k/s_0$ is well-defined.

As an example of the above algorithm, consider $g(z) = 1+4z+2z^2$, then $\vec s =(1,4,2)$. From the first recursion, we obtain $b_2 = 2$ and $\vec s = (-3,-4)$. From the second recursion, we have $b_1=\frac{4}{3}$ and $\vec s = \left(\frac{7}{3}\right)$. As we will now see, the values for $b_1$ and $b_2$ directly give us the ground state as a quantum circuit.

In Section~\ref{sec:circuit}, we derive the following, using the same conditions as listed in Result \ref{result1} and the $b_k$ obtained through Algorithm \ref{eq:recursion}:

\begin{shaded}
\begin{result}[Construction of MPS.]\label{result2}
If $|b_k| \neq 1$ for $k=1,\dots,d$, the ground state of Eq.~\eqref{eq:H} can be constructed with $d$ layers of circuits: $M^{(d)} M^{(d-1)}  \cdots  M^{(1)}  |\psi_p \rangle$,
where $|\psi_p \rangle$ is the ground state of the fixed-point Hamiltonian $H_p = \frac{1}{2}\sum_n h_{n,p}$. Each layer is generated by a fixed-point Hamiltonian as follows:
\begin{align}
M^{(k)} = \exp\left(-\beta_k H_{p+k}\right) \quad \textrm{with } \beta_k = \mathrm{arctanh}(b_k). \label{eq:result2}
\end{align}
This circuit can be rewritten as an MPS with the bond dimension claimed in Result \ref{result1}; see Section~\ref{sec:MPS}. \end{result}
\end{shaded}

The gates $M^{(k)}$ appearing in this result are generically \emph{not} unitary (so one still has to normalize the wave function). However, $|b_k| \neq 1$ implies they are invertible. In fact, if $\lvert b_k \rvert < 1$ for $k=1,\dots d-1$ then we can give an especially efficient unitary circuit representation for the MPS whose unit element scales logarithmically with bond dimension---see Section \ref{sec:unitary}.

Note that since $H_k$ is a sum of commuting terms, $M^{(k)}$ can be written as a product $M^{(k)}=\prod_n M^{(k)}_n$ as follows:
    \begin{align}
M^{(k)}_n=1- a_k h_{n,k+p} \quad \textrm{with } a_k = \frac{b_k}{1+\sqrt{1-b_k^2}} \; . \label{eq:M}
\end{align}
This local form leads to the MPS description and will be important in our analysis below\footnote{For definiteness, if $b_k>1$ then we take $\sqrt{1-b_k^2} = i\sqrt{b_k^2-1}$.}. 

If $\lvert b_k \rvert <1$ then $\beta_k \in \mathbb{R}$ and so $M^{(k)}$ can be seen as  an imaginary time evolution generated by the fixed-point Hamiltonian of the phase with winding number $\omega = k+p$. If $|b_k| > 1$ then $M^{(k)}$ can be written as an imaginary time evolution $\exp(-\textrm{arctanh}(1/b_k)H_{p+k})$ followed by a unitary SPT entangler\footnote{This follows from the identity $\textrm{arctanh}(b_k) = \textrm{arctanh}(1/b_k) - i \textrm{sign}(b_k) \pi/2.$ Note that $W_a^\dagger = W_aP$ where $P$ is the fermion parity. Hence for our purposes we can always use $W_a$.} $W_{p+k}$, where $W_a = \exp( i \frac{\pi}{2} H_{a})$. (It is straightforward to then show that $b_k \in \mathbb R$ in Eq.~\eqref{eq:result2} is equivalent to the wavefunction being real, as required for the BDI class.)
While the imaginary time evolution cannot change the winding number, these SPT entanglers permute the fixed-point Hamiltonian as follows: $W^{\vphantom \dagger}_a H^{\vphantom \dagger}_b W_a^\dagger=W_a^\dagger H^{\vphantom \dagger}_b W^{\vphantom \dagger}_a = H^{\vphantom \dagger}_{2a-b}$. Using this identity, we can move all SPT entanglers so that they act on the initial state, shifting it to the fixed-point ground state with the same winding number as the model under consideration. Result 2 can therefore be paraphrased as follows:
\textit{Generic states on the MPS skeleton of translation-invariant BDI models with winding number $\omega$ are equivalent to sequences of imaginary time evolutions with fixed point Hamiltonians $H_k$ applied to the fixed-point ground state of $H_\omega$.}
We specify generic states to exclude cases with $\lvert b_k \rvert =1$, and note that unlike Result  \ref{result2}, these imaginary time evolutions are not necessarily applied in order of increasing range $k$ due to their transformation under the SPT entanglers. Result \ref{result2} implies Result \ref{result1} through a continuity argument, taking into account cases with $\lvert b_k\rvert=1$, given in Section \ref{sec:specialc}.

An alternative understanding of our construction is as follows. In terms of the polynomial $g(z)$, for each step $k$ in Algorithm~\ref{eq:recursion} we can define
\begin{equation}
    g_{k-1}(z)= g_k(z)-b_k z^k g_k(1/z),
    \label{eq:recursiong}
\end{equation}
with $g_d(z)=g(z)$. The coefficients of $g_{k-1}(z)$
are the entries of $\vec{s}$ after the $k$-th step. The algorithm decreases the degree step by step until we have $g_0(z) \propto z^0$. If we consider a sequence of models by $f_k(z) \propto z^p g_k(z)^2$, we derive Result~\ref{result2} by showing that applying $M^{(k)}$ to the ground state of $f_{k-1}(z)$ gives us the ground state of $f_k(z)$.  The ground state of $f_0(z)\propto z^p$ is the fixed-point state $\ket{\psi_p}$.

We note that the finite-depth circuit representation of the ground state in Eq.~\eqref{eq:M} holds for both infinite as well as finite periodic chains (for fermionic and spin chain representations). However, when rewriting this circuit as a translation-invariant MPS in Section~\ref{sec:MPS}, the treatment will be most natural for the case of an infinitely-long chain.

\subsection{Consequences}
Beyond constructing the MPS ground state on the MPS skeleton, the above results have some interesting consequences.

Firstly, we can construct a path of MPS-solvable models between any two gapped phases, labelled by winding numbers $\omega_1$ and $\omega_2$, as follows. Let us first consider the case $\omega_1-\omega_2=2k$ for some $k\in\mathbb{N}$. Then define the path: $f(z) = z^{\omega_2}(z+a)^{2k}$, where $a\in\mathbb{R}$. For $a=0$ we have $f(z) =z^{\omega_1}$, while for $a\rightarrow\infty$ we have $f(z) =z^{\omega_2}$. At $a=1$ we have a phase transition; at that point, using the results of Section \ref{sec:MCP}, the ground state is an MPS with the fixed-point ground state of $f(z) = z^{\omega_2+k}$. If $\omega_1-\omega_2=2k+1$ then first take $f(z) = z^{\omega_1-2} (z+a)^2$ for $0\leq a \leq 1$. At the point $a=1$ we are connected to the following path (at the point $A=1$): $f(z) = z^{\omega_1-1} (Az+2+A/z)=z^{\omega_1-1}h(z)$. Then taking $A\rightarrow0$ we have a path (with constant ground state independent of $h(z)$) connecting to $f(z)=z^{\omega_1-1}$. We then can use the previous path to connect to $z^{\omega_2}$. This construction is simply an example, and we will encounter other paths in the next section.

Secondly, one can show that \emph{any model} in the BDI class arises as a limit of a sequence of models with an MPS ground state. Indeed, this follows from being able to obtain a generic polynomial as a limit of polynomials which are squares. We showcase this phenomenon explicitly in Section~\ref{sec:ising} by constructing a path of MPS-solvable models which converge toward the quantum Ising chain in a transverse field. We demonstrate how this can be used to extract the scaling dimension $\Delta=1/8$ associated to the Ising universality class \cite{diFrancesco99}. The general claim is proved in the concurrent work Ref.~\cite{JV}.

Thirdly, if we are in the case where we have a unitary circuit representation for the MPS, then we can use this representation to derive a formula for the (string) order parameter. For $p=0$ and $\lvert b_k\rvert<1$, such a unitary representation exists and we obtain:
    \begin{align}
\lim_{N\rightarrow \infty} \lvert\langle Z_1 \dots Z_N \rangle\rvert = \prod_{k=1}^d (1-b_k^2)^k. \label{eq:orderparameterresult}
\end{align}
This result is derived in Section \ref{sec:orderparameter} and is noteworthy since it does not rely on Wick's theorem or the Toeplitz determinant theory that appears in standard approaches (see, for example, \cite{Wu66,Barouch71,Deift2013}; Ref.~\cite{Jones2019} gives results in the notation of this paper for the general BDI class). In fact, using Toeplitz determinant theory on the MPS skeleton leads to a number of interesting exact results---this is explored in the concurrent work \cite{JV}. 

Finally, in Section~\ref{sec:U(1)}, we apply our results to particle-number-conserving models protected by a sublattice symmetry (class AIII), containing deformations of the Su-Schrieffer-Heeger (SSH) chain \cite{Su79}.

\section{Examples\label{sec:example}}
Here we will discuss three examples for which our results can be applied. As a first example we take the simplest case for Result \ref{result1}. This leads us to a model introduced by M. Wolf et al.~\cite{Wolf06}.

The second example introduces a two-parameter model, and we find several MPS-solvable paths that make up the MPS skeleton. 
Certain special cases appear where our results do not strictly apply, however, we can still find the ground state wave functions (these special cases are analyzed in Sections~\ref{sec:MCP} and~\ref{sec:specialc}).

In a third example we discuss how the quantum Ising chain can be approximated by a series of MPS-solvable parent Hamiltonians. This is illustrative of how any model within the BDI class can arise as a sequence of Hamiltonians with an exact MPS ground state.

\subsection{Transition from \texorpdfstring{$\omega=0$ to $\omega=2$}{winding number zero to winding number two}}\label{sec: example 1}

Based on Result \ref{result1}, the simplest example of an MPS-solvable model in the BDI class that one might come up with is one for which $f(z)$ is the square of a first order polynomial (i.e, $d=1$, $p=0$). Let us take:
\begin{align}
    g(z) = (1-\lambda) + \lambda z,
    \label{eq:ex_g_ex1}
\end{align}
with the parameter $\lambda\in\mathbb{R}$. This gives the polynomial
\begin{align}
    f(z) = g(z)^2 = (1-\lambda)^2 + 2\lambda(1-\lambda) z + \lambda^2 z^2
    \label{eq:ex_gsquared_ex1}
\end{align}
that parameterizes a family of models within the space of the three-parameter Hamiltonian
\begin{align}
    H &= t_0 H_0 + t_1 H_1 + t_2 H_2   \label{eq:three_param_ham} \\
    &= \frac{1}{2} \sum_n \left( t_0 Z_n - t_1 X_n X_{n+1} - t_2 X_{n-1} Z_n X_{n+1} \right), \notag
\end{align}
where we have written out the fixed-point models $H_\alpha$ defined in Section~\ref{sec:model}. Note that this model does not have the $\mathbb Z_2 \times \mathbb Z_2$ symmetry that conventionally protects the cluster SPT phase, but it has the anti-unitary symmetry $\left( \prod_n Z_n \right) K$ (where $K$ is complex conjugation) which also protects the cluster model \cite{Verresen17}.

This parameterized family of Hamiltonians is the same path of Hamiltonians that is introduced by M. Wolf et al. in Ref.~\cite{Wolf06} as an example for a quantum phase transition within the MPS-framework, and was later used by A. Smith et al. in Ref.~\cite{Smith19} to simulate a quantum phase transition on a noisy intermediate-scale quantum computer\footnote{The two models are the same under a $\frac{\pi}{2}$-rotation about the $y$-axis and the identification $g = 2\lambda-1$.}.

\begin{figure}[t]
	\centering
	\includegraphics{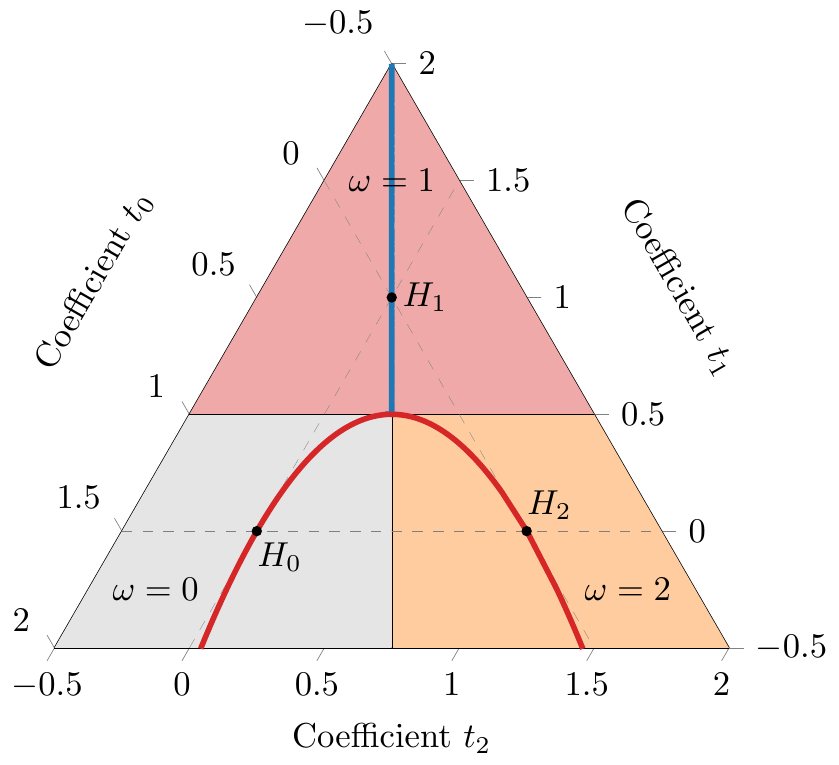}
	\caption{\textbf{MPS skeleton of the Ising-cluster model.} The phase diagram of the model described by the Hamiltonian in Eq.~\eqref{eq:three_param_ham};  each dashed, gray line corresponds to setting one $t_{\alpha}$ to zero. The differently shaded regions show the different phases labelled by the topological invariant $\omega$. In the fermionic representation, these correspond to Kitaev chains with distinct winding numbers. In the dual spin chain formulations, these phases are the trivial paramagnet ($\omega=0$), the Ising magnet ($\omega=1$) and the symmetry-protected topological cluster phase ($\omega=2$). The solid red and blue lines show the parameterized paths along which we can find the MPS representation of the ground state, see Eqs.~\eqref{eq:ex_gsquared_ex1} and~\eqref{eq:ex_zh(z)}, the dashed gray lines show the lines where one parameter equals zero. See also Refs.~\cite{Wolf06,Smith19}.}
	\label{fig:phase_diagram_ex1}
\end{figure}

A global, positive prefactor in front of the Hamiltonian in Eq.~\eqref{eq:three_param_ham} does not change its ground state, so we can normalize the Hamiltonian such that $t_0+t_1+t_2=1$. This makes it possible to draw the phase diagram, as shown in Fig.~\ref{fig:phase_diagram_ex1}. The regions with different colors in the phase diagram indicate the different phases of the model, labelled by the topological invariant $\omega$. 

The solid red line in the phase diagram shows the Hamiltonians belonging to the parameterization in Eq.~\eqref{eq:ex_gsquared_ex1}. With the chosen parameterization, running through $\lambda$ from $-\infty$ to $\infty$ means traversing the red curve from left to right. For $\lambda=0$ the corresponding Hamiltonian is $H_0$, at $\lambda=\frac{1}{2}$ the phase transition occurs and at $\lambda=1$ the corresponding Hamiltonian is $H_2$. Note that this is Kramers-Wannier dual to the disorder-line of the XY chain \cite{Barouch71,Verresen17}.

Within the phase diagram, there is actually another line that corresponds to MPS-solvable Hamiltonians, shown in blue. On this line the polynomial describing the Hamiltonian is
\begin{align}
    f(z) &= \tilde{\lambda} z^2+(1-2\tilde{\lambda} ) z + \tilde{\lambda} \nonumber\\
    &= z\left(\tilde{\lambda} z +(1-2\tilde{\lambda}) +\tilde{\lambda}/z\right)=zh(z)  \label{eq:ex_zh(z)}
\end{align}
with $\tilde\lambda \leq 1/4$. Note that $h(1/z)=h(z)$, has no zeros on the unit circle and has a positive constant term for all $\tilde\lambda <1/4$---hence, we have that the ground state is the same as $f(z)=z$ along this entire path (see the discussion following Result \ref{result1}). For $\tilde{\lambda}=1/4$ we are at the multicritical point ($\lambda=1/2$ on the red curve), while for $\tilde\lambda > 1/4$ we are on a critical line. On this critical line, we still have the form $f(z)=zh(z)$ but now $h(z)$ has zeros on the unit circle and the low energy physics is described by a conformal field theory (CFT); in particular, an SPT-entangled XX model \cite{Sachdev}.

Turning back to the question of finding the MPS representation of the ground state of the model in Eq.~\eqref{eq:ex_gsquared_ex1}, we can apply Result~\ref{result2}. For $b_1$ we simply find ${b_1 =\frac{s_1}{s_0} = \frac{\lambda}{1-\lambda}}$, which gives ${a_1 = \frac{\lambda}{1-\lambda+\sqrt{1-2\lambda}}}$. Note that for $\lambda=\frac{1}{2}$ and $\lambda\rightarrow\pm\infty$ we have that $b_1=\pm1$ and so Result~\ref{result2} does not apply. These special points---which also happen to be the phase transitions---will be discussed as special cases in Section~\ref{sec:MCP}.

\begin{figure}[t]
	\centering
	\includegraphics{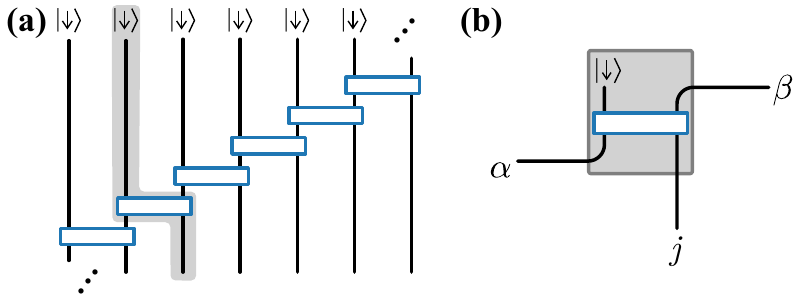}
	\caption{\textbf{Circuit and MPS equivalence for the skeleton of the Ising-cluster model.} \textbf{(a)} Circuit construction for the ground state in the example from $\omega=0$ to $2$, given in Eq.~\eqref{eq: example 1 GS}. The circuit elements $M^{(1)}_n=1-a_1h_{n,1}$ are represented by blue boxes coupling two neighbouring spins (black lines). The repeated unit element is highlighted in gray. \textbf{(b)} The repeating unit element of the circuit forms a tensor $A^j_{\alpha \beta}$, as defined in Eq.~\eqref{eq: example 1 tensor}, which shows the equivalence of the circuit to a matrix product state with bond dimension $\chi=2$.}
	\label{fig:circuit example 1}
\end{figure}

From the circuit construction of the state
\begin{align}\label{eq: example 1 GS}
    |\textrm{gs}\rangle \propto \prod_n M_n^{(1)}|\psi_0\rangle = \prod_n \left(1-a_1 h_{n,1}\right)|\psi_0\rangle
\end{align}
we can then obtain the usual MPS tensors (see Section \ref{sec:MPS} for the definition). The circuit construction and equivalence to an MPS is illustrated in Fig.~\ref{fig:circuit example 1}.

We get the MPS tensor by interpreting the circuit gate as a four-legged tensor, where one ingoing leg acts on a spin $\ket{\downarrow}$, one outgoing one corresponds to the physical index, and the two legs connecting the ladder structure can be interpreted as the virtual legs---this is illustrated in Fig.~\ref{fig:circuit example 1}. Therefore we have
\begin{equation}\label{eq: example 1 tensor}
    A^j_{\alpha\beta} = \bra{\alpha} \bra{j} M_n^{(1)} \ket{\downarrow} \ket{\beta},
\end{equation}
which in this case gives the MPS tensor
\begin{equation}
\begin{aligned}
    A^{\uparrow} &= \left(\begin{array}{cc}
        0 & a_1 \\
        1 & 0
    \end{array}\right) \\
    A^{\downarrow} &= \left(\begin{array}{cc}
        a_1 & 0 \\
        0 & 1
    \end{array}\right).
\end{aligned}\label{eq:Atensor}
\end{equation}
We can compare our solution to the MPS given in Ref.~\cite{Wolf06} for this path. After rotating into the basis of Eq.~\eqref{eq:Atensor} and inserting $g = 2\lambda-1$, the MPS from Ref.~\cite{Wolf06} is
\begin{equation}
\begin{aligned}
    M^{\uparrow} &= \frac{1}{\sqrt{2}}\left(\begin{array}{cc}
        -1 & 2\lambda-1 \\
        \phantom{-}1 & 1 \\
    \end{array}\right) \\
    M^{\downarrow} &= \frac{1}{\sqrt{2}}\left(\begin{array}{cc}
        \phantom{-}1 & 1-2\lambda \\
        \phantom{-}1 & 1 \\
    \end{array}\right).
\end{aligned}
\end{equation}
It can be checked that the matrix
\begin{equation}
    V = \left(\begin{array}{cc}
        \lambda\sqrt{2\lambda-1} & \sqrt{2\lambda-1} \left(\lambda - 1 + i s \sqrt{2\lambda-1}\right) \\
        i s \lambda & \sqrt{2\lambda-1} - i s (\lambda-1) \\
    \end{array}\right),
\end{equation}
where $s$ denotes the sign of $2\lambda-1$, relates the two MPS tensors as $A^j \propto V^{-1} M^j V$. Therefore the two MPS representations are equivalent.

For all values of $\lambda$, we can use the results of Section \ref{sec:unitary} to find a unitary circuit representation. This means that, using Eq.~\eqref{eq:orderparameterresult}, we have the following expression for the order parameter. For the $\omega=0$ phase ($\lambda<1/2$) we have:
\begin{align}
\lim_{N\rightarrow \infty} \lvert\langle Z_1 \dots Z_N\rangle\rvert = 
\frac{1-2\lambda}{(1-\lambda)^2}
\end{align}
and for the $\omega=2$ phase ($\lambda>1/2$):

\begin{align}
\lim_{N\rightarrow \infty} \lvert\langle X_1Y_2\; \prod_{j=3}^N Z_j \; Y_{N+1}X_{N+2} \rangle\rvert  = 
\frac{2\lambda-1}{\lambda^2};
\end{align}
for further details see Section \ref{sec:orderparameter}.
\subsection{Transitions between  \texorpdfstring{$\omega=0$, $\omega=2$ and $\omega=4$}{winding numbers zero, two and four}}

As a second example, we take a look at a path containing a phase transition between phases with the topological invariant $\omega=0$ and $\omega=2$, as well as a transition between phases with $\omega=2$ and $\omega=4$. We note that in the spin representation, $H_0$, $H_2$ and $H_4$ are all in distinct interacting SPT phases protected by the $\mathbb Z_2 \times \mathbb Z_2^T$ symmetry generated by $P = \prod_n Z_n$ and complex conjugation $T=K$ \cite{Verresen17}.

\begin{figure}[t]
    \centering
    \includegraphics[width=0.46\textwidth]{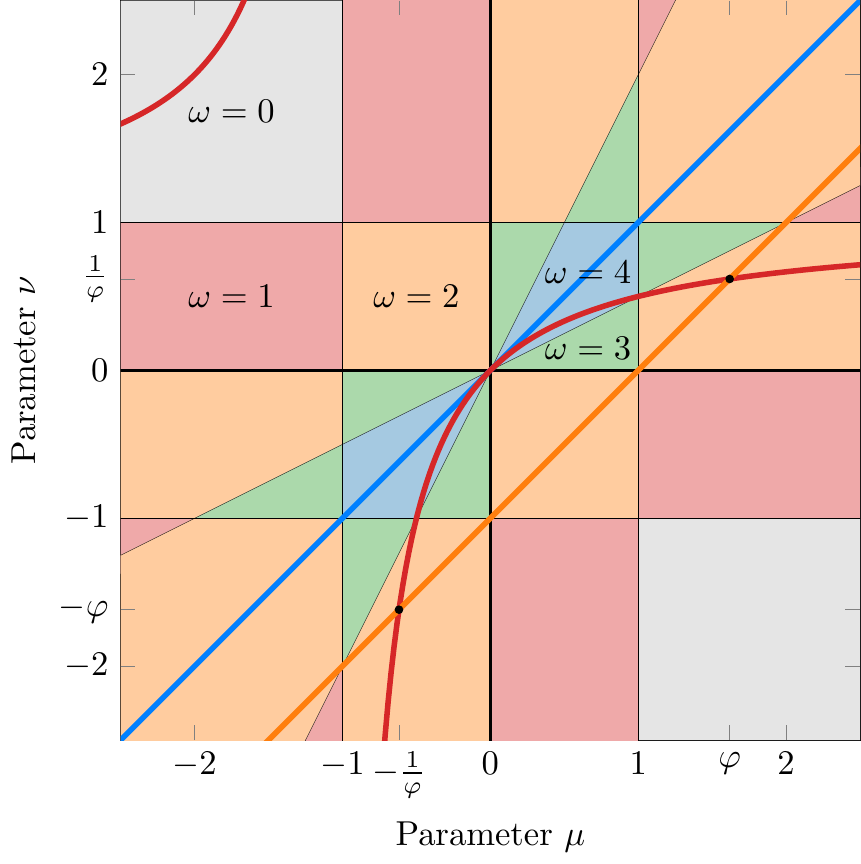}
    \caption{\textbf{MPS skeleton of generalized cluster models or Kitaev chains.} The phase diagram of the model described by the function $f(z)$ in Eq.~\eqref{eq:ex_f}; the differently shaded regions show the different phases labelled by $\omega$. The solid blue, red and orange lines show the parameterized paths along which we can find the MPS representation of the ground state, the blue line corresponds to the case $\nu=\mu$, see Eq.~\eqref{eq:ex_gsquared2}, the red line to the case $\nu=\frac{\mu}{\mu+1}$, see Eq.~\eqref{eq:ex_gsquared}, and the orange line to the case $\nu=\mu-1$, see Eq~\eqref{eq:ex_z2h(z)}. The red and orange lines intersect at the golden ratio $\varphi = \frac{1+\sqrt{5}}{2}$.}
    \label{fig:phase_diagram}
\end{figure}

Let us consider the model described by the polynomial 
\begin{equation}
    f(z) = (z-\mu)(z-\nu)\left(z-\frac{\mu-\nu}{\mu}\right)\left(z-\frac{\mu-\nu}{\nu}\right),
    \label{eq:ex_f}
\end{equation}
with $\mu,\nu\in\mathbb{R}$ to ensure hermiticity of the Hamiltonian.\\
By varying the parameters $\mu$ and $\nu$ we can explore the different phases of the model. The phase diagram is shown in Fig.~\ref{fig:phase_diagram}. There, the differently colored regions correspond to the different phases, which are labelled by the topological invariant $\omega$.

There are two choices for $\nu$ in Eq.~\eqref{eq:ex_f} for which we can express $f(z)$ as the square of a function $g(z)$; if we choose $\nu=\mu$ we find
\begin{equation}
    f(z) = z^2 g(z)^2 = z^2 (z-\mu)^2,
    \label{eq:ex_gsquared2}
\end{equation}
and if we choose $\nu=\frac{\mu}{\mu+1}$ we find
\begin{equation}
    f(z) = g(z)^2 = (z-\mu)^2\left(z-\frac{\mu}{\mu+1}\right)^2.
    \label{eq:ex_gsquared}
\end{equation}
For these particular choices of $\nu$ we can then apply Result~\ref{result2} to find the MPS representation of the ground state. The two options are also plotted as lines in the phase diagram in Fig.~\ref{fig:phase_diagram}, where the blue line indicates the first case and the red line indicates the second case.

There is a third line, shown in the phase diagram in orange, that corresponds to a family of MPS-solvable Hamiltonians, but is not a square of a function $g(z)$. If we choose $\nu=\mu-1$, we find
\begin{align}
    f(z) &= (z-\mu)\big(z-(\mu-1)\big)\left(z-\frac{1}{\mu}\right)\left(z-\frac{1}{\mu-1}\right) \nonumber \\
    &= z^2 \left(z-\left(\frac{\mu^2+1}{\mu}\right)+\frac{1}{z}\right) \left(z-\left(\frac{\nu^2+1}{\nu}\right)+\frac{1}{z}\right) \nonumber \\
    &= \pm z^2 \; h(z).
    \label{eq:ex_z2h(z)}
\end{align}
Here $h(z)$ is a Laurent polynomial that satisfies $h(z)=h(1/z)$, has a positive constant term and no roots on the unit circle for $\mu\notin\lbrace-1, 0, 1, 2\rbrace$. Note that for $0<\mu<1$ we need to factor out a minus sign, so that the constant term of $h(z)$ is positive. This means that the ground state along the orange line in Fig.~\ref{fig:phase_diagram} is that of $H_2$ for $\mu<0$ and $\mu>1$, and that of $-H_2$ for $0<\mu<1$.

Let us now return to the two cases of $f(z)=g(z)^2$ in Eqs.~\eqref{eq:ex_gsquared2} and~\eqref{eq:ex_gsquared}. As the first case (the blue line in Fig.~\ref{fig:phase_diagram}) is essentially the previous example up to replacing $p=0$ by $p=2$, we will focus on the second case here---this is the red line in Fig.~\ref{fig:phase_diagram}. 

In order to apply Result~\ref{result2}, we first need to expand the function $g(z)$ in Eq.~\eqref{eq:ex_gsquared} and calculate the set of $b_k$. Expanded, $g(z)$ becomes
\begin{equation}
    g(z) = \frac{\mu^2}{\mu+1}-\left(\frac{\mu}{\mu+1}+\mu\right) z+z^2.
    \label{eq:ex_g}
\end{equation}
Using Algorithm~\ref{eq:recursion} we can calculate ${b_2=\frac{\mu+1}{\mu^2}}$ and ${b_1 = -\frac{\mu (\mu+2)}{\mu^2+\mu+1}}$, which fully specify the gates $M^{(2)}_n$ and $M^{(1)}_n$.

Finally, we can write the ground state of the model in Eq.~\eqref{eq:ex_gsquared} as
\begin{equation}
    |\textrm{gs}\rangle \propto \left(\prod_n M_n^{(2)}\right) \left(\prod_m M_m^{(1)}\right) |\psi_0\rangle,
\end{equation}
with $|\psi_0\rangle$ being the ground state of $H_0$.

This result holds as long as $|b_k|\neq1$ holds for $k=1,2$ and for $\mu\in\mathbb{R}$. One can check that the cases where this does not hold are $\mu\in\left\lbrace-1, -\frac{1}{2}, 1\right\rbrace$ where $|b_1|=1$, and $\mu=\frac{1}{2}\left(1\pm\sqrt{5}\right)$ where $|b_2|=1$. The values of $\mu$ where $|b_1|=1$ happen to be the phase transitions (see Fig.~\ref{fig:phase_diagram}). At these points the usual procedure fails but it turns out that we can find the gates that construct the ground state in all cases. We make some general points about these exceptional cases in Sections~\ref{sec:MCP} and~\ref{sec:specialc}. In particular, the ground state of the model with $\mu\in\left\lbrace-1, -\frac{1}{2}, 1\right\rbrace$ is constructed in Section~\ref{sec:MCP} and the ground state of the model with $\mu=\frac{1}{2}\left(1\pm\sqrt{5}\right)$ is constructed in Section~\ref{sec:specialc}. While the case $\mu=\frac{1}{2}\left(1\pm\sqrt{5}\right)$ is actually already included in the discussion of the model in Eq.~\eqref{eq:ex_z2h(z)}, Section~\ref{sec:specialc} gives a more general discussion of cases where $|b_k|=1$.

\subsection{A path of MPS for the quantum Ising chain}\label{sec:ising}
We now show how our results can also tell us something about general models in this class. We will take \begin{align}
H_{\mathrm{Ising}} = H_0 + J H_1 =\frac{1}{2}\sum_n\left( Z_n -JX_nX_{n+1}\right)\label{eq:Ising} 
\end{align} as an instructive example of such a model that does not have an exact MPS ground state (for $J \neq 0$). The fermionic chain with this Hamiltonian interpolates between the trivial and the Kitaev chain (with critical point at $|J|=1$); while the corresponding spin chain is the transverse field Ising model. Even though this model is not MPS-solvable, we will construct a sequence of MPS-solvable models which converges towards it. Note that the idea used here to find the path can be generalized to give a path of MPS-solvable models converging towards any Hamiltonian in the BDI class---the proof can be found in Ref.~\cite{JV}. 

The related Laurent polynomial is $f(z)= 1 + Jz$, from which we can read off the topological invariant $\omega =0$ for $|J|<1$ and $\omega=1$ for $|J|>1$. We will focus on the $\omega =0$ phase\footnote{The case $|J|>1$ is Kramers-Wannier dual to $|J|<1$ \cite{Kramers41}, so our results can be applied also to that case.}, as well as the critical point. Of course, $1 + Jz$ is not a square, and thus does not have an exact MPS ground state. However, we can write $f(z) = g(z)^2$ with $g(z) = \sqrt{1+Jz}$. We can then use the series expansion 
\begin{align}
    \sqrt{1+x} =\sum_{n=0}^\infty \left( \begin{array}{c} 1/2 \\ n \end{array} \right) x^n \qquad \textrm{if } |x| \leq 1 \label{eq:square-root}
\end{align}
to expand $g(z)$. This expansion is valid if it converges to $f(z)$ on the unit circle---indeed it is on the unit circle where we connect $f(z)$ to our Hamiltonian (as discussed in Section \ref{sec:results}, the absolute value of $f(e^{ik})$ gives the energy spectrum and, moreover, its phase encodes the single-particle modes \cite{Verresen17}). Hence, if $|J|\leq1$, we can define
\begin{equation}
f_m(z) = g_m(z)^2 \quad \textrm{with } g_m(z) = \sum_{n=0}^m \left( \begin{array}{c} 1/2 \\ n \end{array} \right) J^n z^n. \label{eq:MPSpath}
\end{equation}
This converges to the quantum Ising chain: $\lim_{m\to \infty} f_m(z) = f(z)$. Each $f_m(z)$ corresponds to a Hamiltonian on the MPS skeleton, and has an exact MPS ground state with bond dimension $\chi = 2^m$. Note that this path can be used even for $|J|=1$: for all $m$, all roots of $f_m(z)$ lie strictly outside the unit disk\footnote{This can be shown using Rouch\'{e}'s theorem \cite{JV}.}; hence, $f_m(z)$ gives a path of gapped Hamiltonians that approximate a critical Hamiltonian. 

More explicitly, for any $m \in \mathbb N_+$, the perturbed Ising chain (with $|J| \leq 1$) which has an exact MPS ground state with $\chi=2^m$ is given by
\begin{equation}
H =\frac{1}{2} \sum_n \left( Z_n - J X_n X_{n+1} \right) + J^{m+1} \; \delta H. \label{eq:pathIsing}
\end{equation}
The perturbation $\delta H$ is obtained by calculating the coefficients of $f_m(z) = g_m(z)^2$ and using binomial identities to simplify the double sum, resulting in
\begin{equation}
\delta H =\sum_{\beta = 0}^{m-1} \frac{ (\beta-m)  \left( \begin{array}{c}- \frac{1}{2} \\ \beta  \end{array} \right) \left( \begin{array}{c} -\frac{1}{2} \\ m \end{array} \right)  }{(\beta+m)(\beta+m+1)} J^\beta \; H_{\beta+m+1}  \label{eq:deltaH}
\end{equation}
where $H_\alpha = -\frac{1}{2}\sum_n X_n Z_{n+1} \cdots Z_{n+\alpha-1} X_{n+\alpha}$ are (fixed-point) generalized cluster models.

One can prove that the MPS path in Eq.~\eqref{eq:MPSpath} is the optimal path of MPS approximations in the space of polynomials $g_m(z)$ which do not have roots inside the unit disk; see Appendix~\ref{app:MPSpath} for a proof. Moreover, the same derivation also tells us that the energy density $E_m = \bra{\varphi_m} (Z_n - J X_n X_{n+1})\ket{\varphi_m}/2$, where $ \ket{\varphi_m}$ is the ground state of $f_m(z)$, is given by:
\begin{equation}
E_m = - \frac{1}{2} \sum_{n=0}^{m} \left( \begin{array}{c} 1/2 \\ n \end{array} \right)^2 J^{2n}. \label{eq:MPS_E}
\end{equation}
Indeed, this converges to $E_\infty := \lim_{m\to \infty} E_m$ which is equal to $-\frac{1}{2\pi} \int_{0}^{2\pi} \sqrt{1+J^2+2J\cos k} \mathrm dk$, the ground state energy density of the quantum Ising chain. From Eq.~\eqref{eq:MPS_E} we also learn that the deviation for a given truncation $m$ is
\begin{equation}
E_{m} - E_{\infty} =  \frac{1}{2}\sum_{n=m+1}^\infty  \left( \begin{array}{c} 1/2 \\ n \end{array} \right)^2 J^{2n} < J^{2m+1} |E_\infty|.
\end{equation} 
In particular, for $|J|<1$, the energy deviation decreases exponentially in $m$. Do note that since $m = \log_2 \chi$, this is only a polynomial decay in $\chi$. E.g., if $J=0.5$, then $\Delta E \sim 1/\chi^2$.

We thus obtain a path of MPS with ever-increasing bond dimension that converges to the ground state of the quantum Ising chain. Moreover, the above mechanism (using series expansions) can be applied to any generic model in the BDI class; this is worked out in more detail in the concurrent work \cite{JV} where it is used to analytically derive results about generic models. Note that sequences of free-fermion MPS converging to the ground state of the XY model are investigated in Ref.~\cite{Rams_2015}. The approach there is valid for a particular region of the phase diagram that includes the quantum Ising model; however, in contrast to our path, performing the truncation requires numerical calculations.

\begin{figure}
        \begin{tikzpicture}
   \node at (0,0) {\includegraphics[scale=0.45]{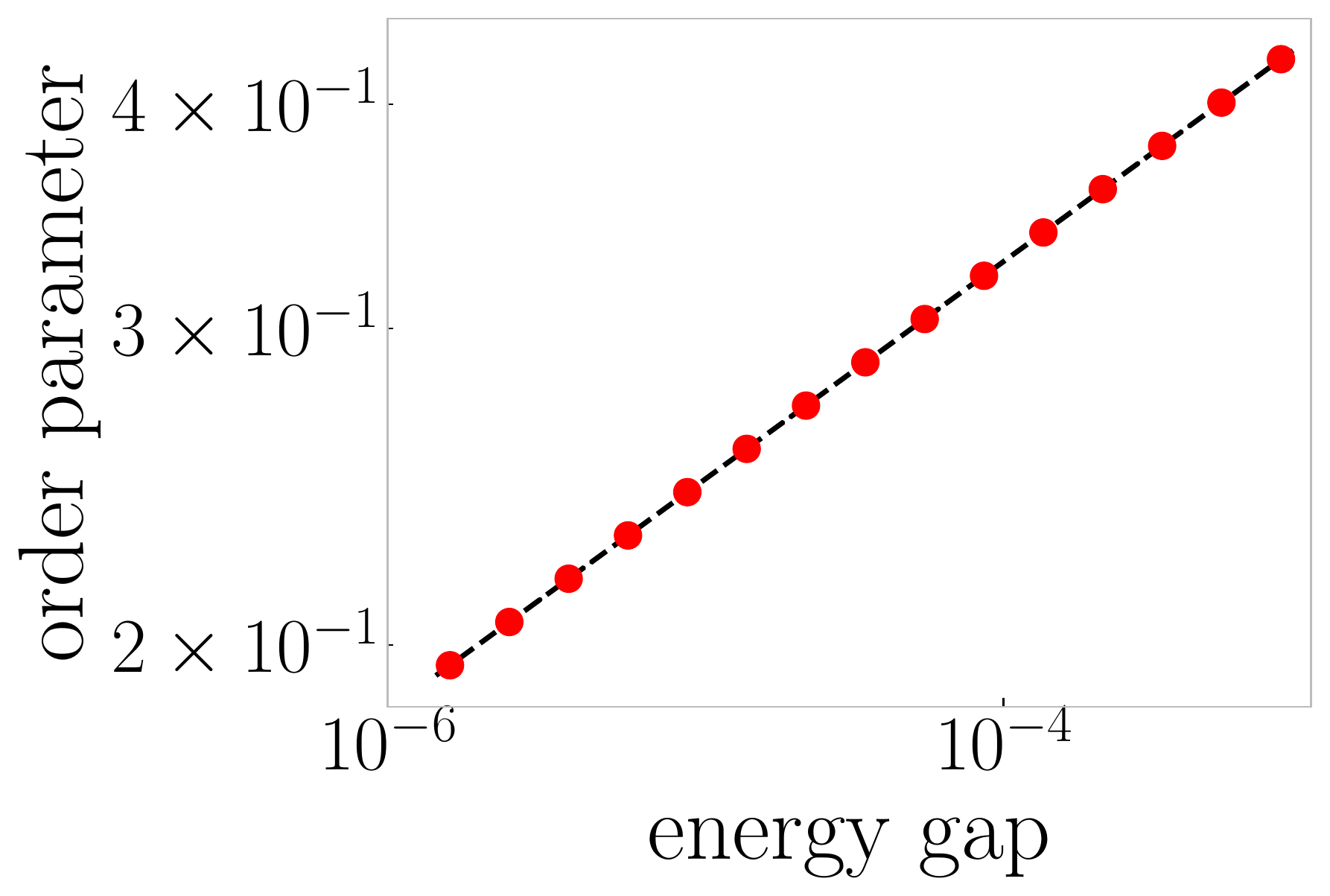}};
      \node at (.75,-2.75) {\large{energy gap}};
      \node[rotate=90] at (-4.25,.25) {\large{order parameter}};
    \end{tikzpicture}
    \caption{\textbf{Scaling along MPS-solvable path that converges to the critical Ising chain.} Red dots correspond to different choices of $m\in\mathbb{N}$ ($400\leq m \leq 2\times 10^5$) labeling the path of MPS-solvable models in Eq.~\eqref{eq:pathIsing}, for which we calculate the energy gap and string order parameter. Both quantities converge to zero at the Ising critical point; the dashed black line gives scaling exponent $\Delta \approx 1/8$ (see Eq.~\eqref{eq:Deltafit}).}
    \label{fig:pathIsing}
\end{figure}

Given this path of MPS ground states approximating the Ising model, it is interesting to see what we can derive about the critical model this way. Let us recall that the Ising CFT has two non-identity local scaling operators $\epsilon$ and $\sigma$ (corresponding to the energy term and order parameter, respectively) as well as two nonlocal ones, $\mu$ and $\psi$ (corresponding to the disorder operator and fermion, respectively) \cite{diFrancesco99}. On the lattice, $\cdots Z_{n-2} Z_{n-1} Z_n \sim \mu(x)$. We will now show how to extract the scaling dimension $\Delta_\mu$ of $\mu$ using the above path of MPS. (Note that due to Kramers-Wannier duality, this also immediately gives us $\Delta_\sigma = \Delta_\mu$.)

We will use the path in Eq.~\eqref{eq:pathIsing} (with $J=1$) where $\delta H$ in Eq.~\eqref{eq:deltaH} detunes it into the trivial paramagnetic phase for any finite $m \in \mathbb N$. This detuning gives a finite energy gap $\varepsilon_\pi$ (at $k=\pi$) and long-range order to the disorder parameter, $\lim_{N\to \infty}\langle Z_1 Z_2 \cdots Z_{N-1} Z_N \rangle =: \langle \mu \rangle^2 \neq 0$. We will obtain both quantities. From their relative scaling $\langle \mu \rangle \sim \varepsilon_\pi^{\Delta_\mu/\Delta_\epsilon}$ and noting that it is a simple argument\footnote{For the quantum Ising chain, $f(z) = 1+ Jz$, hence the gap scales as $\varepsilon_\pi \sim |1-J|$, which thus vanishes linearly as one approaches $J\to 1$.} to derive that $\Delta_\epsilon=1$, we thus extract $\Delta_\mu$.

First, at $J=1$, the gap is given by
\begin{equation}
\varepsilon_\pi = |f_m(e^{i\pi})| = 4  (m+1)^2 \left( \begin{array}{c} 1/2 \\ m+1 \end{array} \right)^2 \sim \frac{1}{\pi m}. \label{eq:gap}
\end{equation}
Second, in Section \ref{sec:orderparameter}, we derive a formula for the order parameter in the $\omega=0$ phase that is applicable to our path, from which we obtain $\langle \mu \rangle = \prod_{k=1}^m \left(1-b_{k}^2 \right)^{k/2}$. Note that $\{b_{k}\}_{k=1,\cdots,m}$ (which implicitly depend on $m$) can be efficiently obtained from the coefficients of $g_m(z)$ by order $m$ multiplications and additions\footnote{This is different to other formulas in terms of roots \cite{JV}, which in general can only be approximated numerically.}. We plot the result in Fig.~\ref{fig:pathIsing} where we find $\langle \mu \rangle \sim 1/{m^{1/8}}$. More precisely, by fitting the exponent (black dashed line), we find the critical exponent associated to the disorder operator of the Ising CFT:
\begin{equation}
    \Delta_\mu = 0.12500004 \pm 5 \times 10^{-8}. \label{eq:Deltafit}
\end{equation}
This agrees with the exact result $\Delta_\mu = 1/8$ \cite{diFrancesco99}. While these exponents are well-known and have alternative lattice derivations, our method gives a path of exact MPS that converges to this critical state.
We mention that $\langle \mu \rangle^2$ can be written as a determinant, and one way to find the scaling dimension is to use Toeplitz determinant theory \cite{Deift2013}. Indeed, using those analytic methods one can obtain the exact asymptotics of the $\mu(x)$ two-point function in the ground state of $H_{\mathrm{Ising}}$, giving the scaling dimension above. We note that Ref.~\cite{Hartwig1969} gives a sequence of approximations to this determinant, based on expanding the square-root as in Eq.~\eqref{eq:square-root}. We point out that our treatment does not require this theory. 

An advantage of the above is that it gives an analytic expression for a path of MPS-solvable parent Hamiltonians which is optimal in some respect, as explained above. However, it is not optimal in the space of all MPS-solvable Hamiltonians: by allowing roots of $g_m(z)$ inside the unit disk, the variational energy can be decreased. As a $\chi=2$ example, consider $f_\textrm{var}(z) = \frac{1}{z^2} ( z - z_1)^2 \left(z-\mathcal{Z}_1\right)^2$ where $|z_1|<1$ and $|\mathcal{Z}_1|>1$, which has winding number $\omega = 0$. For any such choice of roots, the resulting MPS will have $\chi=2$. One can optimize these roots to minimize the variational energy with respect to the quantum Ising chain Hamiltonian given above. The variational energy is given by the negative real root of greatest absolute value of the equation:
\begin{align}
\mathcal{E}^4-4 \mathcal{E}^3 +\left(2 J^2+\frac{9}{2}\right) \mathcal{E}^2+12 J^2 \mathcal{E}+ J^4+ \frac{9J^2}{2}=\frac{27}{16};\label{eq:venergy}
\end{align}
see Appendix \ref{app:energy} for details.

For $J=0$, Eq.~\eqref{eq:venergy} gives the exact ground state energy density, and the deviation from the exact result increases with $0\leq \vert J\rvert \leq 1$. For the critical quantum Ising chain (i.e., $\vert J\rvert =1$), the best variational energy for $\chi=2$ (using $f_\textrm{var}(z)$) is
\begin{equation}
\frac{1}{2}\left( 3 + \frac{4}{\sqrt[3]{3} \lambda} - \frac{2\lambda}{\sqrt[3]{9}} \right) \bigg|_{\lambda = \sqrt[3]{63+11\sqrt{33}}} \approx -0.6349, \label{eq:variationalMPS}
\end{equation}
which can be compared to the exact result $-\frac{2}{\pi} \approx  -0.63662$. We thus see that the $\chi=2$ free-fermion MPS can reproduce the correct energy density within $0.3 \%$.
For $\lvert J \rvert=1$, this is almost an order of magnitude better than the variational energy given by Eq.~\eqref{eq:MPS_E} for $m=1$, namely $-5/8 = -0.625$, which is within $2\%$ of the true energy density. 

\section{Analysis}\label{sec:analysis}
To derive the results stated in Section~\ref{sec:results}, we first show that the ground state is frustration-free if the associated polynomial is of the form $f(z)=z^pg(z)^2$. In Section \ref{sec:circuit}, using Witten's conjugation method \cite{Witten1982,Wouters2020}, we then derive an explicit (non-unitary) circuit mapping its ground state to a fixed point wavefunction. This circuit can then be explicitly rewritten as a matrix product state with the claimed bond dimension, this is derived in Section \ref{sec:MPS}.  

The results in Section~\ref{sec:results} were stated for both the fermionic chain and the Jordan-Wigner dual spin chain simultaneously; however, 
certain sections below are more straightforwardly presented with one or the other picture. In particular, Sections~\ref{sec:frustration}-\ref{sec:circuit} make use of the fermionic notation for ease of presentation. 
In Section \ref{sec:MPS} we explain how to construct the MPS tensor. While fermionic MPS are well understood \cite{Bultinck2017,Cirac2020}, the spin chain representation is more conventional. Despite working with the Hilbert space of the spin chain, it will still be useful to present certain formulas using fermionic operators throughout this work. The underlying Jordan-Wigner transformation is given by:
\begin{align}
\gamma_n = \bigg(\prod_{j<n} Z_j\bigg) X_n,  \qquad
\tilde\gamma_n= \bigg(\prod_{j<n} Z_j \bigg) Y_n .\label{eq:JW}
\end{align}
While care usually has to be taken about the precise meaning of the product over sites $j<n$, dependent on boundary conditions, in Section \ref{sec:MPS} we will implicitly be working in case of an infinitely-long chain, such that these subtleties do not arise. The Jordan-Wigner dual expressions for $h_{n,\alpha}=i\tilde\gamma_n\gamma_{n+\alpha}$ were already given in Eq.~\eqref{eq:hnspin}, which includes the identity $Z_n=i \tilde\gamma_n\gamma_n$. We also have:
\begin{align}
\gamma_n\gamma_{n+\alpha}&= -i Y_n Z_{n+1}\cdots Z_{n+\alpha-1}X_{n+\alpha}\nonumber\\
\tilde\gamma_n\tilde\gamma_{n+\alpha}&= i X_n Z_{n+1}\cdots Z_{n+\alpha-1}Y_{n+\alpha}.\label{eq:JW2}
\end{align}
The ground state of $H_0$ is denoted by $\ket{\psi_0}$. This corresponds to the completely filled fermionic state $c^\dagger_nc_n\ket{\psi_0}=\ket{\psi_0}$; while for spins the corresponding state is $\ket{\downarrow\cdots\downarrow}$.  Using these identities, all formulas below can be transformed from fermions to spins and vice-versa.

\subsection{The Hamiltonians are frustration-free \label{sec:frustration}}

A frustration-free model is one where the Hamiltonian can be written as a sum of terms such that each term is individually minimized in the ground state \cite{Perez2007,Wouters2020}. Here we derive the frustration-free property of the above systems. This will also form the starting point of the wavefunction construction in Section~\ref{sec:circuit}. 

If $f(z) = z^p g(z)^2$ with $g(z) = \sum s_\alpha z^\alpha$, then Eq.~\eqref{eq:H} can be written as
\begin{equation}
H = \frac{i}{2} \sum_n \left( \sum_\alpha s_\alpha \tilde \gamma_{n-\alpha} \right)  \left( \sum_\beta s_\beta \gamma_{n+\beta+p} \right) \label{eq:frustration}
\end{equation}
which is confirmed by expanding it out. Note that each term indexed by $n$ in Eq.~\eqref{eq:frustration} has eigenvalues $\pm |\vec s|^2$. We now show that the ground state minimizes \emph{each} term in Eq.~\eqref{eq:frustration}, i.e., that the energy density is $e_0 = -\frac{1}{2}|\vec s|^2$. This is the defining property of a frustration-free model.

For any $f(z)$, the ground state energy density can be expressed as
\begin{equation}
    e_0 = - \frac{1}{2\pi } \int \frac{\varepsilon_k}{2} \mathrm d k = -\frac{1}{4\pi i}\oint \frac{\sqrt{f(z)f(1/z)}}{z} \mathrm dz
\end{equation}
where the contour integral is along the unit circle. If $f(z) = z^p g(z)^2$, this simplifies to
\begin{equation}
e_0 = -\frac{1}{4\pi i} \oint \frac{g(z)g(1/z)}{z} \mathrm dz = -\frac{1}{2}\sum_k s_k^2 = -\frac{1}{2}|\vec s|^2. \label{eq:energy}
\end{equation}
As explained above, this shows that the model is frustration-free.

For what follows, it will prove to be useful to define $\Gamma_n = \sum_{\alpha} s_\alpha \big(\gamma_{n+\alpha+p} - i \tilde \gamma_{n-\alpha} \big)$ and then to rewrite Eq.~\eqref{eq:frustration} as
\begin{align}
H = \frac{1}{4} \sum_n\left( \Gamma_n^\dagger \Gamma_n -2{\lvert s\rvert^2}\right).\label{eq:frustration2}
\end{align}
By expanding Eq.~\eqref{eq:frustration2} and observing that terms of the form $i\gamma_n\gamma_m$ and $i\tilde \gamma_n \tilde \gamma_m$ do not survive (either by explicit computation or by noting the complex-conjugation symmetry of the model), one verifies that it equals Eq.~\eqref{eq:frustration}. Similarly, one sees that the frustration-free property of Eq.~\eqref{eq:frustration} is equivalent to the ground state $|\psi\rangle$ of Eq.~\eqref{eq:frustration2} satisfying $\Gamma_n^\dagger \Gamma_n|\psi\rangle = 0$ for all $n$. For the fermionic case under consideration\footnote{Note that the Jordan-Wigner transformation of Eq.~\eqref{eq:frustration2} for periodic boundary conditions will give nonlocal `boundary' terms involving phase factors. However, using the translation-invariance of the state, a frustration-free local spin Hamiltonian is obtained by simply dropping the nonlocal phase factor.}, this is in turn equivalent\footnote{E.g., writing the singular value decomposition $\Gamma_n = U SV$, we see that $\Gamma_n^\dagger \Gamma_n = V^\dagger S^2 V$. Hence, they have the same zero eigenvalues/eigenvectors.} to $\Gamma_n |\psi \rangle = 0$.

\subsection{Constructing the circuit (Result \ref{result2})} \label{sec:circuit}

The MPS parent Hamiltonian construction leads to a frustration-free Hamiltonian where a given MPS is the ground state. The converse typically holds, although not all frustration-free models have MPS ground states\footnote{References \cite{Matsui1998,Matsui2013,Ogata2016} do show that under certain additional conditions, frustration-free Hamiltonians have MPS ground states.}: an example is given in  Ref.~\cite{Bravyi2012}. Here we give a direct proof that this is the case for translation invariant BDI models with $f(z) = z^p g(z)^2$ by explicitly deriving the quantum circuit that constructs the ground state.  

Firstly, note that it is sufficient to prove this for $p=0$. Indeed, one can shift $f(z) \to z^q f(z)$ by shifting all $h_{n,\alpha} \to h_{n,\alpha+q}$ (see Eqs.~\eqref{eq:H} and \eqref{eq:fz}). Hence, we see that starting from the $p=0$ result one recovers $p\neq 0$ by shifting $h_{n,\alpha} \to h_{n,\alpha+p}$ in all formulas.

Secondly, we assume a positive overall sign of $f(z)$.  Different global signs of $f(z)$ are related by the unitary transformation $S$, given by \begin{align}
S= \prod_n i \tilde{\gamma}_{2n-1}\tilde{\gamma}_{2n}.\label{eq:sigmatoggle}
\end{align}  
This global sign generically does not affect the analysis so we can account for it by applying $S$ to the final state. (The only exceptions to this are cases with zeros on the unit circle, discussed in Section \ref{sec:MCP}.) Under conjugation by $S$ we invert the gates $M^{(k)}_n$; i.e., let $M^{(k)}_n =1 - A_k h_{n,k}$, then $SM^{(k)}_n S^{\dagger}=1 + A_k h_{n,k}\propto (M^{(k)}_n)^{-1}$.

To reach Result \ref{result2}, we derive the following stronger result:
\begin{shaded}
\begin{result}[Relating circuits and polynomials.]\label{result3}
If $|\psi_{\mathrm{i}} \rangle$ is the initial ground state associated to some polynomial $f_{\mathrm{i}}(z) =  g_{\mathrm{i}}(z)^2$, then for any $k\in \mathbb Z$ and $B_k \in \mathbb R$ with $\lvert B_k \rvert \neq 1$, the transformed state
\begin{align}
\ket{\psi_{\mathrm{f}}} := \exp\left( - \mathrm{arctanh}(B_k) \; H_k \right) \ket{\psi_{\mathrm{i}}}, \label{eq:psinew}
\end{align}
is the ground state for $f_{\mathrm{f}}(z) =  g_{\mathrm{f}}(z)^2$ where 
\begin{align}
g_{\mathrm{f}} (z) = g_{\mathrm{i}}(z) + B_k z^k g_{\mathrm{i}}(1/z). \label{eq:gnew}
\end{align}\end{result}
\end{shaded}
Note that $ \exp\left( - \mathrm{arctanh}(B_k) \; H_k \right) =M^{(k)}= \prod_n M_n^{(k)}$ where
\begin{align}
M^{(k)}_n=1- \color{black}A_k h_{n,k} \quad \textrm{for} ~A_k = \frac{B_k}{1+\sqrt{1-B_k^2}} .\label{eq:MAB}
\end{align}

\subsubsection{Analysis: {Construction of MPS}.} \label{sec:analysisMPSconstruction}
Using Result~\ref{result3}, one can start from the trivial case $g(z) = 1$ (with trivial ground state $\ket{\psi_0}$) and successively apply layers of gates to obtain the ground state of the desired $g(z)$. 
Let us first take an example. Starting with $g_{\mathrm{i}}(z)=1$ and applying a layer generated by the Kitaev or Ising chain, $M^{(1)}$, we obtain $g_{\mathrm{f}}(z) = 1+B_1z$. With a second layer, $M^{(2)}$, applied to $g_{\mathrm{i}}(z)=1+B_1z$, we obtain the ground state of $g_{\mathrm{f}}(z) = 1+B_1z + B_2z^2(1+B_1/z) = B_2 z^2+B_1(1+B_2)z+1$. If we set $B_1 = \frac{4}{3}$ and $B_2=2$, we recover the polynomial $g(z) = 2z^2 + 4z+1$ that we discussed in Section~\ref{sec:results}. 

More generally, let us show that the recursion in Algorithm~\ref{eq:recursion} leads to the desired ground state. Define $b_k$ for $k=1,\dots,d$ using this recursion. Recall that we start from the Hamiltonian corresponding to $f(z) = g_d(z)^2$ and then the recursion defining $b_k$ amounts to Eq.~\eqref{eq:recursiong}, which is, for each $k$, given by:
\begin{align}
g_{k-1}(z) = g_k(z) -b_k z^k g_k(1/z).
\end{align}
This is equivalent to:
\begin{align}
 (1-b_k^2) g_k(z) =\left(g_{k-1}(z) +b_k z^kg_{k-1}(1/z)\right).    
\end{align}
As long as $\lvert b_k \rvert \neq 1$ we can thus invert the recursion; the term $(1-b_k^2)\neq 0$ is an unimportant constant. Hence, setting $B_k= b_k$, we can work up from the ground state of $f_0(z)= 1$ to $f(z) = g_d(z)^2$ using Result \ref{result3}. 

We now have Result~\ref{result2} for all values of $p$, i.e., the ground state of $f(z) = z^p g_d(z)^2$ is given as a product 
\begin{align}\label{eq: layered product}
    \ket{\psi} = M^{(d)} M^{(d-1)}\cdots M^{(1)} \ket{\psi_p}
\end{align}
as long as $\lvert b_k\rvert \neq 1$ for any $k$. We now note that the fixed-point wave function $|\psi_p\rangle$ can itself be written down as a circuit, the precise form depending on whether $p$ is odd or even:
\begin{equation}
\begin{array}{lllll}
\ket{\psi_{2q}} &= W_q\ket{\psi_0}  \\
\ket{\psi_{2q+1}}&= W_{q+1}\ket{\psi_1} &= W_{q+1} \prod_n P_n \ket{\psi_0}
\end{array} \label{eq:entangler}
\end{equation}
where $W_q$ are the SPT entanglers introduced above and
\begin{align}P_n = \frac{1}{\sqrt{2}}(1 - h_{n,1}) \label{eq:projector}\end{align} 
are commuting projectors. Due to the two cases in Eq.~\eqref{eq:entangler}, it is helpful to define $q\in\mathbb{Z}$ and $r\in \{0,1\}$ such that $p=2q+r$. By writing the SPT entangler $W_q$ as a product of unitary and mutually commuting operators \begin{align}
W_q = \prod_n \frac{1}{\sqrt{2}}(1 - i h_{n, q}),
\end{align} we have the following representation for the fixed-point wave function $|\psi_p\rangle$:
\begin{equation}\label{eq:psip}
\begin{aligned}
    |\psi_{2q}\rangle &= \left(\prod_n U_n \right) |\psi_0 \rangle,  \\
    |\psi_{2q+1} \rangle &= \left(\prod_n U_n \right) \left(\prod_n P_n \right) | \psi_0 \rangle ,
\end{aligned}
\end{equation} where $U_n =  \frac{1}{\sqrt{2}}(1 - i h_{n, q+r})= \frac{1}{\sqrt{2}}(1 - i h_{n, \lceil p/2 \rceil})$. For notational convenience we have suppressed the dependence on $q$. Then combining Eq.~\eqref{eq: layered product} with Eq.~\eqref{eq:psip} we have a circuit construction of the ground state starting from $\ket{\psi_0}$. 

Note that Result~\ref{result3} is the fundamental statement, and can be used to transform between ground states of any two Hamiltonians in our class that are related by Eq.~\eqref{eq:gnew} (including relating Hamiltonians that are cases with $\lvert b_k\rvert =1$ and where we do not have a construction using Result~\ref{result2}); for a particular choice of transformations we derive Result~\ref{result2} from Result~\ref{result3}.

\subsubsection{Analysis: {Relating circuits and polynomials}}\label{sec:analysisResult3}
To derive Result~\ref{result3}, we start with a ground state $\ket{\psi_\mathrm{i}}$ corresponding to a polynomial $f_\mathrm{i}(z) = g_{\mathrm{i}}(z)^2$. Writing $g_{\mathrm{i}}(z) = \sum_\alpha s_\alpha z^\alpha$, we know that $\ket{\psi_\mathrm{i}} $ is annihilated by $\Gamma_n$ in Eq.~\eqref{eq:frustration2} (with $p=0$). Hence, $M^{(k)}\ket{\psi_\mathrm{i}}$ is annihilated by $\tilde \Gamma_n := M^{(k)} \Gamma_n {M^{(k)}}^{-1}$, implying that it is the ground state of $H= \frac{1}{2} \sum_n \tilde \Gamma_n^\dagger \tilde \Gamma_n$. 
All that remains is to calculate $\tilde \Gamma_n$ to confirm that it corresponds to the polynomial defined in Eq.~\eqref{eq:gnew}.

Since $M_n^{(k)}= 1-A_k i \tilde \gamma_n \gamma_{n+k}$ (remember that we have set $p=0$), its inverse is well-defined because $A_k^2 \neq 1$ (equivalently, $|B_k| \neq 1$, see Eq.~\eqref{eq:MAB}). Hence, up to an irrelevant global factor, we have
\begin{align}
M^{(k)} \gamma_n {M^{(k)}}^{-1} &\propto (1-A_k i \tilde\gamma_{n-k} \gamma_n)\gamma_n ( 1+A_k i \tilde \gamma_{n-k} \gamma_n) \nonumber\\
&= (1+A_k^2) \gamma_n -2 A_k i \tilde \gamma_{n-k} \nonumber \\
&\propto \gamma_n - B_k i \tilde \gamma_{n-k},
\end{align}
where in the last step we divided by $(1+A_k^2)$.
Similarly,
\begin{equation}
    M^{(k)} \tilde \gamma_n {M^{(k)}}^{-1} \propto \tilde \gamma_n + B_k i  \gamma_{n+k}.
\end{equation}
Taken together, we see that $\Gamma_n = \sum_\alpha s_\alpha(\gamma_{n+\alpha} - i \tilde \gamma_{n-\alpha})$ gets mapped to
\begin{align}
\tilde \Gamma_n &= \sum_\alpha s_\alpha (  \gamma_{n+\alpha} + B_k \gamma_{n-\alpha+k} - i (  \tilde \gamma_{n-\alpha} + B_k \tilde \gamma_{n+\alpha-k} ) )\nonumber \\
&= \sum_\alpha \left( s_\alpha + B_k s_{k-\alpha}\right) (\gamma_{n+\alpha} - i \tilde \gamma_{n-\alpha}).
\end{align}
Hence, the transformed state is the ground state corresponding to $f(z) = g_{\mathrm{f}}(z)^2$ with
\begin{align}
g_{\mathrm{f}}(z) = \sum_\alpha \left( s_\alpha + B_k s_{k-\alpha}\right) z^\alpha = g_{\mathrm{i}}(z) + B_k z^kg_{\mathrm{i}}(1/z).
\end{align}

This completes the proof. Note that if instead $B_k = \pm 1$, the gate in Eq.~\eqref{eq:MAB} becomes a projector onto the ground state of $\pm H_k$. Hence in that case, $\ket{\psi_{\mathrm{f}}}$ is zero if $\ket{\psi_\mathrm{i}}$ is the ground state of $f(z) = \mp z^k$, otherwise $\ket{\psi_{\mathrm{f}}}$ in Eq.~\eqref{eq:psinew} is the ground state of $f(z) = \pm z^k$. See Section~\ref{sec:specialc} for further discussion of these cases.

We note that this way of constructing frustration-free Hamiltonians---i.e., where $\Gamma_n$ is conjugated by an invertible operator $M$---is known as the Witten conjugation method \cite{Witten1982,Wouters2020}. Usually, one simply chooses $M$ and considers the resulting frustration-free Hamiltonian. What is special to our case is that we have an explicit formula for a set of $M$ that are quadratic fermionic gates and can be used to generate any frustration-free model in the BDI class.
\subsection{MPS representation \label{sec:MPS}}
\begin{figure}[t!]
\includegraphics[width=\columnwidth]{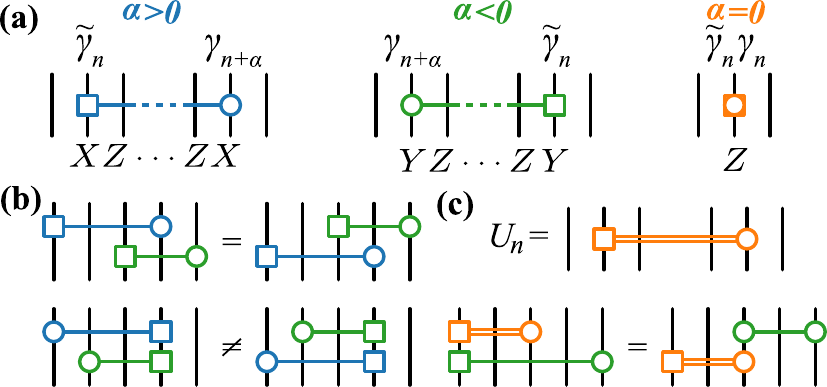}
\caption{\textbf{Graphical notation for the circuit construction of the ground states.} \textbf{(a)} Definition of graphical notation for gates of the form $(1+ch_{n,\alpha})$ with different values of $\alpha$. The squares indicate $\tilde{\gamma}$ Majorana operators while circles correspond to $\gamma$ operators. These gates can be written in terms of spin string operators using Eq.~\eqref{eq:hnspin}. \textbf{(b)} Commutation relations for the gates. These gates commute unless symbols of the same type are on the same wire. \textbf{(c)} Graphical notation for the unitary gates ${U}_n$ defined in Eq.~\eqref{eq:psip}, with double lines to distinguish them from the general form. When one of the end symbols lines up with that of another gate they satisfy the relation shown graphically, which is written explicitly in Eq.~\eqref{eq: U commutation} and in Appendix~\ref{app:MPS}.}\label{fig: graphical notation} 
\end{figure}

Given the explicit circuit construction of the ground state from the previous section, namely Eqs.~\eqref{eq: layered product} and~\eqref{eq:psip}, we will now show that this corresponds exactly to a finite bond dimension translation-invariant MPS. This means that the ground state can be written as the contraction of a translation-invariant tensor network, where the virtual indices between unit cells have finite dimension. For a spin chain, a translation invariant MPS, with MPS tensor $A$, is a state of the form
\begin{align}
\ket{\psi} = \sum_{j_1,\dots j_N}\tr(A^{j_1}A^{j_2}\dots A^{j_N})\ket{j_1 \dots j_N}.
\end{align}
For fixed $j$, $A^j$ is a $\chi \times \chi$ matrix, where $\chi$ is the bond dimension \cite{Cirac2020}. The fermionic case is similar, for details see Ref.~\cite{Bultinck2017}.
For the purposes of defining the MPS tensor, $A$, it is simplest to contract an index of the circuit with the state $\ket{\downarrow}_n$, and so we will work in the spin chain picture. The algebraic steps involved apply (by definition) in the same way to spin or fermionic operators; and for ease of presentation we will continue to use fermionic notation for the gates. These fermionic operators can be interpreted as short-hand for the spin operators as given in Eqs.~\eqref{eq:hnspin} and \eqref{eq:JW2}.

\begin{figure}[tb!]
\includegraphics[width=\columnwidth]{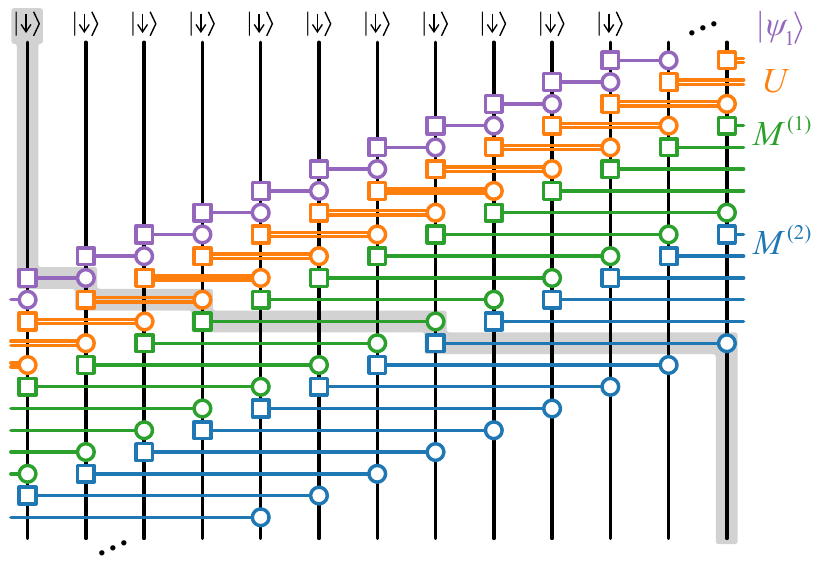}
\caption{\textbf{Example of the mapping from circuit to MPS for $d=2, p=3$ using the graphical notation in Fig.~\ref{fig: graphical notation}.} The different colored gates correspond to the different layers in the state construction. The gray box indicates the repeating circuit element equivalent to an MPS tensor. }\label{fig: circuit to MPS 1}
\end{figure}

To better understand the product of operators in Eq.~\eqref{eq: layered product}, we introduce a graphical notation, defined in Fig.~\ref{fig: graphical notation}. All of the operators appearing in this product have the same form, namely, $1+c h_{n,\alpha} = 1 + ic\tilde{\gamma}_n \gamma_{n+\alpha}$ for some $c\in \mathbb{C}$.
We represent the sites by black ``wires'', similar to quantum circuit diagrams. The operators are then denoted by a colored ``gate'' with ends labelled with either circles denoting $\gamma_n$ operators, or squares for $\tilde{\gamma}_n$ operators. Using this notation, an example of a product of the form in Eq.~\eqref{eq: layered product} is shown in Fig.~\ref{fig: circuit to MPS 1}. 

\begin{figure}[tb!]
\includegraphics[width=\columnwidth]{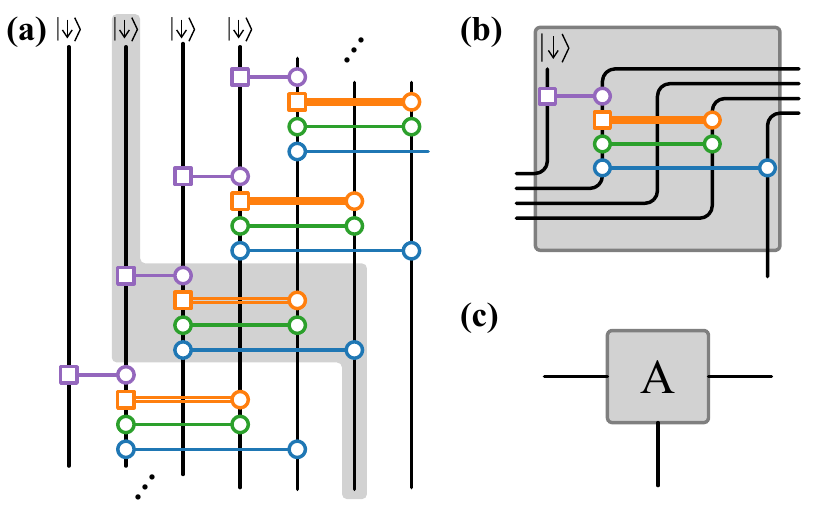}
\caption{\textbf{Example of the mapping from circuit to MPS for $d=2, p=3$ after simplifying.} (See Appendix \ref{app:MPS}) \textbf{(a)} corresponds to Fig.~\ref{fig: circuit to MPS 1} after using the commutation relations for the gates. \textbf{(b)} shows the repeating elements of the sequential circuit. \textbf{(c)} shows the equivalent MPS tensor. }\label{fig: circuit to MPS 2}
\end{figure}

The equivalence to MPS is established by grouping the gates in this product into a repeating unit element, illustrated by the gray box in Fig.~\ref{fig: circuit to MPS 1}. This repeating element has one wire that corresponds to a site and several wires that connect to different unit cells, corresponding to the virtual indices of an MPS tensor. The bond dimension of the corresponding MPS is $\chi=2^N$, where $N$ is the number of wires connecting the unit cells, see Figs.~\ref{fig: circuit to MPS 2}(b-c).

While we have put the circuit in MPS form, this provides a very loose bound for the bond dimension $\chi = 2^N$, where $N$ is quadratic in $d$ and linear in $p$. However, by using the commutation properties of the gates we can provide a much tighter bound on the bond dimension---in fact, we conjecture that this bound gives the optimal bond dimension, and this can be proved in certain cases. These gates are able to commute past each other except when the same symbol in our graphical notation is acting on the same spin, i.e., we cannot bring a circle past a circle or a square past a square, see Fig.~\ref{fig: graphical notation}(b). Furthermore, we are able to bring the unitary gate $U_n$ (defined in Eq.~\eqref{eq:psip}) past those appearing in $M^{(k)}$, which results in the algebraic relation shown in Fig.~\ref{fig: graphical notation}(c). Let us consider here $p=2q+r>0$, (see Appendix \ref{app:MPS} for the more general expression), then
\begin{align}
    M^{(k)}_{n} U_{n} = U_{n} \tilde{M}^{(k)}_{n+q+r},
\end{align}
where  $\tilde{M}^{(k)}_{n+q+r} = 1 -i a_k \gamma_{n+q+r}\gamma_{n+k+p}$.
To see this,
\begin{align}\label{eq: U commutation}
U^{\dagger}_n M^{(k)}_n U_n =& 1 - \frac{a_k}{2}(1+ih_{n,q+r})h_{n,k+p}(1-ih_{n,q+r})\nonumber\\
=& 1 - \frac{a_k}{2}(1+ih_{n,q+r})^2h_{n,k+p}\nonumber\\
=& 1 - i a_k h_{n,q+r}h_{n,k+p}\nonumber\\
=& 1 - ia_k \gamma_{n+q+r}\gamma_{n+k+p}.\end{align} 
 This is shown schematically in Fig.~\ref{fig: graphical notation}. Using Eq.~\eqref{eq:JW2}, we have the corresponding spin operator.

These algebraic relations allow us to drastically reduce the bond dimension, as shown for the particular example of $d=2, p=3$ in Fig.~\ref{fig: circuit to MPS 2}(a). This more compact form follows from the non-trivial application of the commutation relations for the gates and the algebraic identities using the $U_n$ unitary gates and is explained in detail in Appendix~\ref{app:MPS}. In general the bond dimension is given by $\chi$, where
\begin{align}\label{eq: bond dimension bound}
\log_2 \chi = \lceil\textrm{range}(H)/2\rceil.
\end{align}
We show this in Appendix \ref{app:MPS} using the methods introduced here. The bond dimensions for different values of $d$ and $p$ are shown in Tab.~\ref{tab:chi}.

The bond dimension in Eq.~\eqref{eq: bond dimension bound} is an upper bound for the bond dimension required for an exact MPS representation of the ground state. That is, this bond dimension is sufficient for an exact representation. In the case that we have a gapped model on the MPS skeleton with no mutually inverse zeros\footnote{As seen in Section \ref{sec: example 1}, there are cases with mutually inverse roots where this upper bound does not give the optimal bond dimension. These are models defined by $f(z) = \tilde{f}(z) h(z)$ where $h(z) = h(1/z)$, $h(z)$ has a positive constant term and has no zeros on the unit circle. In such cases, we conjecture that the bond dimension will be determined by $\tilde{f}(z)$ according to Eq.~\eqref{eq: bond dimension bound}.}, we believe that this bond dimension is also necessary; i.e., Eq.~\eqref{eq: bond dimension bound} gives the optimal bond dimension. (For gapless points in the MPS skeleton, we show in Section \ref{sec:MCP} that the ground state can be found by considering a related gapped model.)  In Ref.~\cite{JV}, it is explicitly shown in the spin chain representation that when $p=-d$ and $p$ is even, we have a \emph{lower bound} on $\chi$ that coincides with Eq.~\eqref{eq: bond dimension bound}, and hence this proves the optimality in this case. This is proved by analyzing ground state correlation functions in these models\footnote{Denote zeros of $g(z)$ inside the unit circle by $z_j$ and zeros outside the unit circle by $\mathcal{Z}_k$ and let $S=\{z_1,\dots,z_{n_z},\mathcal{Z}_1^{-1},\dots \mathcal{Z}_{n_\mathcal{Z}}^{-1}\}$. The result in \cite{JV} also assumes that given any subset of $S$, if we take the product of zeros in that subset then the absolute value of that product is different to any other subset except for subsets containing any conjugate roots---this condition holds generically for gapped models with $f(z)=z^pg(z)^2$.}. To test the formula for the optimal bond dimension more generally, we compare the analytical upper bound with the bond dimension obtained from finding the ground state numerically using the density matrix renormalization group (DMRG) \cite{White1992a,Hauschild18} with explicitly conserved global $\mathbb{Z}_2$ symmetry. We find that the numerically and analytically obtained bond dimension perfectly coincide for all cases tested, suggesting that this bond dimension is indeed optimal beyond the cases where we have an analytic proof. It would be of interest to see if the methods of Ref.~\cite{Navascues18} could be used to find a lower bound that coincides with our upper bound.

\begin{table}[tb!]
\setlength{\arrayrulewidth}{0.3mm}
\renewcommand{\arraystretch}{1.5}
\centering
\begin{tabular}{|@{}c@{}|@{}c@{}|@{}c@{}|@{}c@{}|@{}c@{}|@{}c@{}|@{}c@{}|@{}c@{}|@{}c@{}|@{}c@{}| }
\hline
\multicolumn{10}{|@{}c@{}|}{\cellcolor{white}$ \chi =2^{\lceil\text{range}(H)/2\rceil}$} \\
\rowcolor{gray!40}
\hline
\hphantom{'}{\textcolor{white}{$d$}\textbackslash $p$}\hphantom{'}& \hphantom{'}$-4$\hphantom{'} &  \hphantom{'}$-3 $\hphantom{'} & \hphantom{'}$-2$\hphantom{'} &  \hphantom{'}$-1 $\hphantom{'} &  \hphantom{-'}$0$\hphantom{-'} &  \hphantom{-'}$1$\hphantom{-'} &  \hphantom{-'}$2$\hphantom{-'} &  \hphantom{-'}$3$\hphantom{-'} & \hphantom{-'}$4$\hphantom{-'}\\
\rowcolor{white}
\hline{}
$\textcolor{white}{1}$\cellcolor{gray!40} & $4$ & $4$ & $2$ & $2$ & $2$ & $4$ & $4$& $8$ & $8$\\
\rowcolor{white}
\hline
$\textcolor{white}{2}$\cellcolor{gray!40} & $4$ & $4$ & $2$ & $4$ & $4$ & $8$ & $8$ & $16$ & $16$\\
\rowcolor{white}
\hline
$\textcolor{white}{3}$\cellcolor{gray!40} & $4$ & $4$ & $4$ & $8$ & $8$ & $16$ & $16$ & $32$ & $32$\\
\rowcolor{white}
\hline
$\textcolor{white}{4}$\cellcolor{gray!40} & $4$ & $8$ & $8$ & $16$ & $16$ & $32$ & $32$ & $64$ & $64$\\
\hline
\end{tabular}
\caption{The bond dimension $\chi$ for $d\in\{1,2,3,4\}$ and ${p\in\{-4,-3,-2,-1,0,1,2,3,4\}}$.}
\label{tab:chi}
\end{table}
\section{Special cases}
\label{sec:special}

\begin{figure*}[!t]
    \centering
    \includegraphics[width=0.85\textwidth]{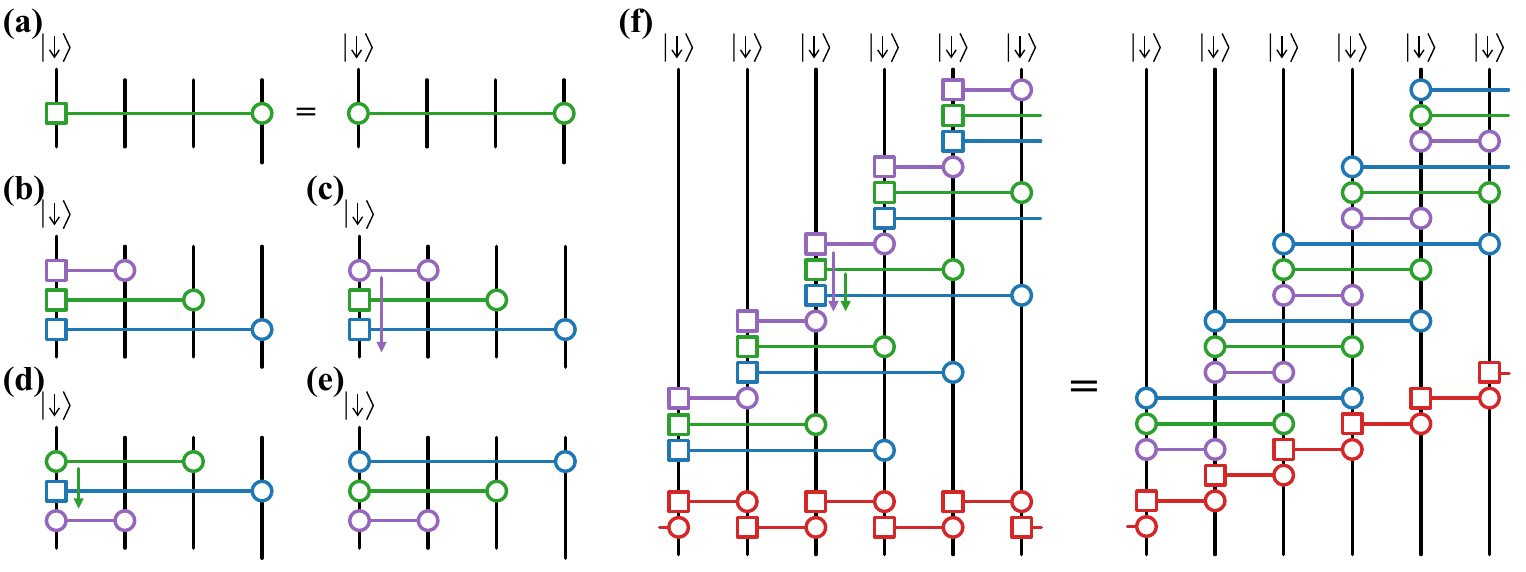}
    \caption{\textbf{Putting a circuit into unitary form.} \textbf{(a)} Illustration of the substitution explained in the main text that allows us to change a square to a circle when acting directly on the initial state $|\!\downarrow\rangle$ (see Eq.~\eqref{eq:sub}). The substitution turns the gates $P_n$ (if $p=1$) and those appearing in $M_n^{(k)}$ into a unitary gate if $a_k\in\mathbb{R}$. \textbf{(b-e)} The steps for repeatedly using the substitution in \textbf{(a)} and commutation relations to bring the circuit into unitary form. The figures shown correspond to either $p=0, d=3$ or to $p=1,d=2$, but can be applied more generally as explained in the main text. \textbf{(f)} Equality between MPS circuit and sequential unitary circuit. Without the unitary SPT entangler $W_{\alpha=1}$ (red), this applies to examples with $p=0, d=3$ or with $p=1, d=2$. With the SPT entangler, it applies to examples with $p=-4, d=3$ or with $p=-3, d=2$.}
    \label{fig: unitary circuit}
\end{figure*}

\subsection{Unitary version}\label{sec:unitary}

The circuit construction described in the previous sections is generally made from non-unitary gates $M^{(k)}$ and projectors $P$. However, in certain cases we can instead use a circuit made entirely of unitary gates. Note that while any MPS can be written as a sequential unitary circuit~\cite{Schon2005}, the depth of the repeating unit element would scale with bond dimension $\chi$, whereas our construction scales as $\log\chi$. Such a unitary circuit representation may be useful for processing quantum states on quantum computing platforms. We first explain how this works for $p=0,1$, and $a_k \in \mathbb{R}$ for all $k$ (equivalently $\lvert b_k\rvert <1$ for all $k$, recall that $a_k$ is defined in Eq.~\eqref{eq:M}). We then show how this can be extended to all $p\in\mathbb{Z}$ and $a_d$. We show in Section \ref{sec:orderparameter} that the condition that $\lvert b_k\rvert <1$ for $k< d$ means that the zeros of $g(z)$ are either all inside or all outside the unit circle.

\subsubsection{Circuit construction}

To put our circuits into unitary form we need to use a substitution that is demonstrated in Fig.~\ref{fig: unitary circuit}(a). Schematically, we are able to turn a square symbol acting directly on the initial state into a circle (see Fig.~\ref{fig: graphical notation} for the definition of these symbols). Explicitly, we have the relation
\begin{equation}
    (1- i a_k \tilde{\gamma}_n \gamma_{n+k}) \mathbb{X}(n) \ket{\psi_0} = (1 - a_k \gamma_n \gamma_{n+k})\mathbb{X}(n) \ket{\psi_0}, \label{eq:sub}
\end{equation}
where $\mathbb{X}(n)$ is any operator that is \emph{not} supported on site $n$.  This follows from $i \tilde \gamma_n \gamma_n \ket{\psi_0} = Z_n  \ket{\downarrow\cdots\downarrow} = -  \ket{\downarrow\cdots\downarrow}$.
On the right-hand side of Eq.~\eqref{eq:sub} we have a gate proportional to 
\begin{align}
\tilde{V}^{(k)}_n &= (1-a_k\gamma_n\gamma_{n+k})/\sqrt{1+a_k^2}
\label{eq:defV}
\end{align} which is unitary for $a_k\in \mathbb{R}$. 

In the case that $p=0$ or $p=1$, from Eq.~\eqref{eq:result2} (and the discussion following it), our circuit is of the form 
\begin{align}
\ket{\psi} = \prod_n (1-i A_{p+d}\tilde{\gamma}_n\gamma_{n+d+p})\cdots \prod_n (1-i A_1\tilde{\gamma}_n\gamma_{n+1})|\psi_0\rangle
\end{align}
for some $A_k \in \mathbb{R}$. Indeed, if $p=0$, then $A_k = a_k$ whereas if $p=1$, then $A_1=1$ (this is the gate $P_n$ in Eq.~\eqref{eq:projector} that projects the initial state into the ground state of $H_1$) and $A_{k>1} = a_{k-1}$.
We can then make the substitution in Eq.~\eqref{eq:sub} for each of the $(1-i A_1\tilde{\gamma}_n\gamma_{n+1})$. The resulting $\tilde{V}_n^{(1)}$ will commute past the other gates and so we can bring it down, as illustrated in Figs.~\ref{fig: unitary circuit}(c-d). 
We then have the gate $(1-i A_2\tilde{\gamma}_n\gamma_{n+2})$ acting on the state $\mathbb{X}(n)\ket{\psi_0}$, where $\mathbb{X}(n)$ is some operator not supported on site $n$. Repeatedly using the substitution and commuting gates, we end up with a version of the circuit consisting of only unitary gates, as shown in Fig.~\ref{fig: unitary circuit}(f).

By applying extra layers of unitary gates we can extend the set of states that we can construct with unitary circuits to include all $p\in\mathbb{Z}$ and no restriction on $b_d$ (i.e. $a_d$ can be real or on the unit circle). This is achieved using the SPT entanglers $W_\alpha$, recall that these are given by:
\begin{equation}\label{eq:pivot}
    W_\alpha = \exp \left( i \frac{\pi}{4} \sum_n h_{n,\alpha} \right) = \prod_n \frac{(1 + i \tilde{\gamma}_n \gamma_{n+\alpha})}{\sqrt{2}}.
\end{equation}
Conjugating by $W_\alpha$ corresponds to the mapping $t_n \rightarrow t_{2\alpha-n}$ in the Hamiltonian in Eq.~\eqref{eq:H}, or, equivalently, in the polynomial $f(z)$ in Eq.~\eqref{eq:fz}.
By using $W_{d+p}$ we can remove the constraint on $a_d$\footnote{Note that applying this transformation takes $g(z)\rightarrow z^dg(1/z)$. This takes $b_d\rightarrow 1/b_d$ and $b_k \rightarrow b_k$ for $k<d$. Hence, we still require $\lvert b_k \rvert <1$ for $k<d$.}. Furthermore, combining two of these entanglers results in an even shift, that is $W_{\beta} W_{\alpha} : t_n \rightarrow t_{n + 2(\beta-\alpha)}$. Starting from $p=0,1$ this allows us to transform to any value of $p$. Since all of the gates appearing in the product in Eq.~\eqref{eq:pivot} commute, we can similarly collect gates into a repeating unit cell. 

It is important to note that this alternative construction of states using $W_\alpha$ does not generally correspond to the optimal bond dimension found in previous sections. However, the unitary circuit representation may be of practical use, for instance in processing these states on a quantum computer~\cite{Smith19}.

\subsubsection{A formula for the order parameter}\label{sec:orderparameter}
As a result of the unitary circuit representation given above, we can derive a formula for the order parameter in those models. In this section we will work in the spin chain picture. Let us consider $g(z) = \sum_{k=0}^d s_k z^k$, such that the corresponding $\lvert b_k \rvert <1$. Then, we have:
\begin{align}
\lim_{N\rightarrow \infty} \lvert\langle Z_1 \dots Z_N \rangle\rvert = \prod_{k=1}^d (1-b_k^2)^k, \label{eq:order}
\end{align}
where the left-hand side is the string order parameter for $\omega =0$.

To prove this, consider the ground-state correlation function for a half-infinite string $\langle \prod_{j=m}^\infty Z_j \rangle$, which is given by:
\begin{align}
&\bra{\downarrow \downarrow \cdots \downarrow } \prod_{n} \tilde U_n    \prod_{j=m}^\infty Z_j \prod_{n} \tilde U_n^\dagger \ket{\downarrow \downarrow \cdots \downarrow } \label{eq:orderparameter}\\
&\tilde U_n= \tilde{V}_n^{(d)} \dots \tilde{V}_n^{(1)} \nonumber.
\end{align}
Note that $\prod_{n} \tilde U_n $ is such that $\tilde U_n$ acts before $\tilde U_{n+1}$ on the string---see also Fig. \ref{fig: unitary circuit}(f).
In terms of spins, we have: \begin{align}\tilde V_n^{(k)}=(1+ i a_kY_n Z_{n+1} \dots Z_{n+k-1} X_{n+k})/\sqrt{1+a_k^2}.\end{align} Then, for $(m-k)\leq n<m$ we have
\begin{align}
   \tilde V_n^{(k)}&  \prod_{j=m}^\infty Z_j \tilde V_n^{{(k)}^\dagger} = (1-b_k^2)^{1/2}\prod_{j=m}^\infty Z_j + \textrm{const}\times \mathcal{O}_{n,k}
 \label{eq:orderparameter2}
\end{align}
where
\begin{align}
\mathcal{O}_{n,k}= Y_n Z_m \dots Z_{m+k-1} Y_{n+k}\prod_{j=n+k+1}^\infty Z_j.
\end{align}
For all other values of $n$, $\tilde V_n^{(k)}$ commutes with  $\prod_{j=m}^\infty Z_j$. Moreover,  $\tilde V_n^{(k')}$ commutes with $\mathcal{O}_{n,k}$ for $k'>k$. One can then see that, since $\bra{\downarrow}_n Y_n \ket{\downarrow}_n=0$, the second term in Eq.~\eqref{eq:orderparameter2} does not contribute. Then by replacing:
\begin{align}
\tilde V_n^{(k)}&  \prod_{j=m}^\infty Z_j \tilde V_n^{{(k)}^\dagger} \rightarrow (1-b_k^2)^{1/2}\prod_{j=m}^\infty Z_j\end{align}
for $\tilde{V}_{m-d}^{(d)},\tilde{V}_{m-d+1}^{(d-1)},\tilde{V}_{m-d+1}^{(d)},\dots \tilde{V}_{m-1}^{(1)}, \dots\tilde{V}_{m-1}^{(d)}$, we reach \begin{align}
\left\langle \prod_{j=m}^\infty Z_j \right\rangle=\prod_{k=1}^d (1-b_k^2)^{k/2},
\end{align}
implying the result.

Now, within the BDI class, $\lim_{N\rightarrow \infty} \lvert\langle Z_1 \dots Z_N \rangle\rvert \neq 0$ is equivalent to being in the gapped phase with $\omega=0$ \cite{Jones2019} (and in particular, $\lim_{N\rightarrow \infty} \lvert\langle Z_1 \dots Z_N \rangle\rvert \rightarrow 0$ implies that the gap closes). This means that, since $p=0$, Eq.~\eqref{eq:order} tells us that if $\lvert b_k \rvert <1$ then all zeros of $g(z)$ are outside the unit circle. In fact these conditions are equivalent: if all zeros of $g(z)$ are outside the unit circle, then all $\lvert b_k \rvert <1$. To see this, consider $g_0(z) = 1+ \epsilon z^d$ for $\epsilon <1$. This has all zeros outside the unit circle, and $b_d=\epsilon$, $b_{k<d}=0$. Let $g(z)$ have all zeros outside the unit circle, we can tune the zeros of $g_0(z)$ to the zeros of $g(z)$ along paths outside the unit circle, and this corresponds to a path of gapped Hamiltonians. Moreover, the $b_k$ vary continuously along this path, and at no point along the path can we have $\lvert b_k \rvert\rightarrow 1$ as this would contradict the fact that the path is gapped, hence, $g(z)$ also has all $\lvert b_k \rvert < 1$. As explained above, the case $\lvert b_d \rvert >1$ and $\lvert b_k \rvert <1$ for $k<d$ can be analyzed by applying the SPT entangler $W_d$. This transformation maps zeros of $g(z)$ to inverse zeros of $g(z)$, and so must correspond to the case that $g(z)$ has all zeros inside the unit circle.

Recall that $g(z) = \sum_{k=0}^d s_k z^k$, and let us fix $s_0=1$ for convenience. As an illustration, we can evaluate Eq.~\eqref{eq:order} for $d=1$:
\begin{align}
\lim_{N\rightarrow \infty} \lvert\langle Z_1 \dots Z_N \rangle\rvert = (1-s_1^2), \label{eq:d=1}
\end{align}
where $\lvert s_1\rvert<1$ and for $d=2$:
\begin{align}
\lim_{N\rightarrow \infty} \lvert\langle Z_1 \dots Z_N \rangle\rvert = \left((1+s_2)^2-s_1^2 \right)(1-s_2)^2, \label{eq:d=2}
\end{align}
where $\lvert s_2 \rvert<1$ and $\lvert s_1 \rvert < \lvert 1+ s_2 \rvert$. If we denote the zeros of $g(z)$, which lie outside the unit circle, by $\mathcal{Z}_1, \dots, \mathcal{Z}_d$, then we also have:
\begin{align}
\lim_{N\rightarrow \infty} \lvert\langle Z_1 \dots Z_N \rangle\rvert = \prod_{k,k'=1}^d (1-\mathcal{Z}^{-1}_k\mathcal{Z}^{-1}_{k'}). \label{eq:order2}
\end{align}
This is a special case\footnote{As explained in Ref.~\cite{JV}, in the case where all zeros of $g(z)$ are outside the unit circle this result follows from the results of Ref.~\cite{Hartwig1969}.} of a more general formula for the order parameter in terms of zeros of $g(z)$, given in Ref.~\cite{JV}. This means that:
\begin{align}
\prod_{k,k'=1}^d (1-\mathcal{Z}^{-1}_k\mathcal{Z}^{-1}_{k'})=\prod_{k=1}^d (1-b_k^2)^{k}.
\end{align}
Note that for $d=1$, $s_1=1/\mathcal{Z}_1$, so using Eq.~\eqref{eq:d=1} we see immediately that this equality is satisfied. For $d=2$ one can show directly that Eq.~\eqref{eq:d=2} and \eqref{eq:order2} are equal. 

Eq.~\eqref{eq:order} applies for the case $p=0$. By applying SPT entanglers, it is an immediate consequence that if we allow general $p\in\mathbb{Z}$ with the condition $\lvert b_k \rvert <1$ then:
\begin{align}
\lim_{N\rightarrow \infty} \lvert\langle \mathcal{O}_p(1)\mathcal{O}_p(N+1) \rangle\rvert = \prod_{k=1}^d (1-b_k^2)^k,
\end{align}
where $\mathcal{O}_p$ is the (string) order parameter for the phase with winding number $p$. In the spin chain representation $\mathcal{O}_p$ is local (non-local) for $p$ odd (even)---general definitions are given, for example, in Ref.~\cite{Jones2019}. We have that $\mathcal{O}_1(n)=X_n$, $\mathcal{O}_{-1}(n)=Y_n$ and $\mathcal{O}_2(n)=(\prod_{j<n}Z_j) Y_nX_{n+1}$.

\subsection{ \texorpdfstring{$U(1)$}{U(1)} symmetric chains}\label{sec:U(1)}

Thus far we have focused on the BDI class which contains (superconducting) pairing terms. The Kitaev chain \cite{Kitaev2001} is the generating SPT of this class, in the sense that all topological phases are obtained by considering stacks. In this section, we discuss the AIII class which preserves particle number and has the SSH chain \cite{Su79} as its generator. We will show that it can be embedded into the BDI class, thereby offering a reinterpretation of some of our results.

Similar to Eq.~\eqref{eq:H}, we can define the following fermionic hopping model:
\begin{align}
H= \sum_{\alpha, n} \left(\tau_\alpha c^\dagger_{B,n} c^{\vphantom \dagger}_{A,n+\alpha} + \overline\tau_\alpha c^\dagger_{A,n} c^{\vphantom \dagger}_{B,n-\alpha} \right) .\label{eq:H U(1)}
\end{align}
As before, we define the range of this Hamiltonian to be the largest $\lvert \alpha \rvert$ of all non-zero $\tau_\alpha$. For $\tau_\alpha \in \mathbb{C}$, this is a general translation-invariant model in the AIII class. Let us first discuss the special case $\tau_\alpha \in \mathbb{R}$. As explained in Appendix \ref{app:U(1)}, Eq.~\eqref{eq:H U(1)} can be rewritten as a translation-invariant Majorana chain where the range has been doubled; more precisely, it has the form of the BDI Hamiltonian in Eq.~\eqref{eq:H} with $t_{2\alpha} = \tau_\alpha$ and $t_{2\alpha-1}=0$. Hence for $\tau_\alpha \in \mathbb{R}$ our results above apply directly to these models. In particular, there exists an MPS with bond dimension $\log_2 \chi = \textrm{range}(H)$; and if all $\lvert b_k \rvert \neq 1$ then we have a construction of this MPS.

To state the analogous construction, let us define the fixed point Hamiltonians \begin{align}
    H^{U(1)}_k =\sum_n\left( c^\dagger_{B,n}c^{\vphantom\dagger}_{A,n+k} +h.c. \right).
\end{align}
For instance, the case $k=1$ is the SSH chain \cite{Su79}.
We define the associated polynomial $f(z) = \sum_{\alpha}\tau_\alpha z^{\alpha} = z^p g(z)^2 $. (As before, the quasiparticle dispersion relation is $\varepsilon_k = |f(e^{ik})|$. Recent work has established the relation between $f(z)$ and topological edge modes \cite{Balabanov21}.) If we calculate the $b_k$ (and corresponding $\beta_k$) exactly as above, then we have that the ground state is given by $\hat M^{(d)} \hat M^{(d-1)}  \cdots  \hat M^{(1)}  |\tilde{\psi}_{\tilde p}  \rangle$ where
\begin{align}
\hat{M}^{(k)}&= \exp\left(-\beta_k H^{U(1)}_k \right).
\end{align}
As before, these $\hat{M}^{(k)}$ correspond to imaginary time evolutions with fixed point Hamiltonians, followed by SPT entanglers for $\lvert b_k \rvert>1$.

In Appendix \ref{app:U(1)} we show that the general Hamiltonian given in Eq.~\eqref{eq:H U(1)} with complex hopping $\tau_\alpha \in \mathbb C$ is equivalent to a Majorana chain with a \emph{two-site unit cell}, for which we have not given an explicit construction of the ground state in the present work. In the concurrent work \cite{JV}, the existence of an exact MPS ground state is proved in the case that $f(z) = \sum_{\alpha}\tau_\alpha z^{\alpha} = z^p g(z)^2 $, but no construction or upper bound on the bond dimension is given. We conjecture that the same bond dimension formula holds in this case (i.e., that $\log_2 \chi = \textrm{range}(H)$).
\subsection{Multicritical points}
\label{sec:MCP}
Since $\lvert f(e^{ik})\rvert$ gives the single-particle energy spectrum, if the polynomial $f(z)$ has zeros on the unit circle, at points $z=e^{i k_n}$, then the system is gapless. Let us consider gapless models on the MPS skeleton: the starting point is, as above, that $f(z) = z^p g(z)^2h(z)$, where $h(z)$ has no zeros on the unit circle, and so any zeros on the unit circle must have even multiplicity. In the phase diagram of translation-invariant BDI models, these are multicritical points. Rather than applying our algorithm immediately, we first simplify the problem by reducing to an equivalent gapped model. This is possible due to the even multiplicity of all zeros on the unit circle. In Appendix \ref{app:multicritical} we show that the ground state of a gapless model given by $f_{\textrm{gl}}(z) =  z^p g_\textrm{gl}(z)^2h(z)$ is the same as the ground state of a closely related gapped model $f_{\textrm{g}}(z) = (-1)^{m_0/2} z^{p+N_c/2} g_{\textrm{g}}(z)^2$.
Here $g_{\textrm{g}}(z)$ is the polynomial $g_\textrm{gl}(z)$ after dividing by $(z-e^{i k_n})$ for all zeros on the unit circle, $N_c/2 \in \mathbb{Z}$ is the number of zeros of $g_\textrm{gl}(z)$ on the unit circle counting multiplicity and $m_0$ is the multiplicity of the zero at $z=1$ (if there is no zero there, $m_0=0$). This means that we can find the ground state at the multicritical point by applying our methods to this related gapped model. 

To see how this works in practice, we can refer back to our earlier examples. In the first example, see Eq.~\eqref{eq:ex_gsquared_ex1}, we have a gapless point at $\lambda=1/2$. In that case $f_{\textrm{gl}}(z) = (z+1)^2$ and this implies ${f}_{\textrm{g}}(z)=z$. Hence, this ground state is simply given by the Ising ferromagnet. Similarly for $\lambda\rightarrow \pm \infty$ (normalizing $f(z)$ appropriately) we have
$f_{\textrm{gl}}(z) = (z-1)^2$, leading to ${f}_{\textrm{g}}(z)=-z$.

In the second example, see Eq.~\eqref{eq:ex_gsquared}, we have gapless points when $\mu\in\{-1/2,\pm 1\}$. Then the ground state can be obtained from
\begin{equation}
{f}_{\textrm{g}}(z) = \begin{cases}  z(z+1/2)^2 &\text{if}{\;}\mu = -1/2 \\ -z(z-1/2)^2 &\text{if}{\;}\mu = 1 \\z&\text{if}{\;}\mu = -1 \end{cases}.
\end{equation}
Note that the case $\mu=-1$ requires taking a well-behaved limit for
the ratio $f_{\textrm{g}}(z)/\lvert f_{\textrm{g}}(z)\rvert$.  
\subsection{Cases where  \texorpdfstring{$\lvert b_k \rvert =1$}{|bk|=1}}\label{sec:specialc}
Above we impose the condition that $\lvert b_k \rvert \neq 1$. This ensures that both the gates $M^{(k)}$ defined in Eq.~\eqref{eq:M} and the recursion in Eq.~\eqref{eq:recursiong} are invertible, and moreover that Algorithm~\ref{eq:recursion} is unambiguous. Here we discuss what happens when we relax this condition. 

First, take a model defined by $f(z)=g(z)^2$ and recall the approach in Section \ref{sec:analysisMPSconstruction}. By inverting the recursion in Eq.~\eqref{eq:recursiong} and then fixing $B_k=b_k$ in Result~\ref{result3}, we could transform a fixed-point ground state into the ground state of $f(z)$. We could equally have set $B_k=-b_k$ in Result~\ref{result3}, giving a transformation from the ground state of $f(z)$ to a fixed-point ground state. These are equivalent since $b_k\rightarrow -b_{k}$ is the same as $M^{(k)}\rightarrow (M^{(k)})^{-1}$.

In the case that $\lvert b_k \rvert \neq 1$ for $k>k_0$ and $\lvert b_k \rvert = 1$ for $k_0$, the second point of view is helpful. We can then use Result~\ref{result3} for each $k>k_0$ to write:
\begin{align}
\ket{\psi'}=(M^{(k_0+1)})^{-1}\dots (M^{(d)})^{-1}\ket{\psi}, \label{eq:psi'}
\end{align}
where $\ket{\psi}$ is the ground state of $f(z)=g(z)^2$ and $\ket{\psi'}$ is the ground state of $f(z)=g_{k_0}(z)^2$.

Our method breaks down at the next step because, as explained in Section \ref{sec:analysisResult3}, if $\lvert b_{k_0}\rvert= 1$ then applying
\begin{align}
\prod_n (1+a_{k_0}h_{n,k_0}) = 
\prod_n (1+b_{k_0}h_{n,k_0})
\end{align}
amounts to applying the projector $P^{(k_0)}$ (note we have a $+b_{k_0}$ here because we set $B_k=-b_k$). 
There is a special case where this can work---when the state $\ket{\psi'}$ that the projector acts on is an eigenstate of the projector.

In particular, our method is still able to construct the ground state if the projector $P^{(k_0)}$ annihilates this state\footnote{If the projector $P^{(k_0)}$ acts as identity, we would also know the state and could then construct the initial state by inverting the other gates. However,
given the construction of Algorithm~\ref{eq:recursion} such a case never arises. See also Appendix~\ref{app:bk=1}.}, i.e.,
\begin{equation}
  P^{(k_0)}\ket{\psi'}=0
  \label{eq:pk0}.
\end{equation}
Due to translation symmetry, we can conclude that $\ket{\psi '}$ is the ground state of $\pm H_{k_0}$. In Appendix~\ref{app:bk=1} we show that this case applies if and only if $g_{k_0-1}(z)=0$ (equivalently: in the application of Algorithm~\ref{eq:recursion} the vector $\vec{s}$ for the iteration step ${k=k_0-1}$ vanishes).
Moreover, we show that given $b_{k_0} 
=\pm 1$,  $\ket{\psi'}$ is the ground state of $b_{k_0}H_{k_0}$, i.e.,  $\ket{\psi'}= S^{(1-b_{k_0})/2}\ket{\psi_{k_0}}$. Thus, from Eq.~\eqref{eq:psi'}, we can construct the ground state of our initial model by:
\begin{align}
\ket{\psi} 
&=M^{(d)} \cdots M^{(k_0+1)}  S^{(1-b_{k_0})/2}\ket{\psi_{k_0}}.
\end{align}
In the above discussion, we took $p=0$. As in Section \ref{sec:circuit}, the result for general $p$ follows by applying SPT entanglers that shift $h_{n,\alpha}\rightarrow h_{n,\alpha+p}$ and $\ket{\psi_{k_0}}\rightarrow \ket{\psi_{k_0+p}}$.

These results are relevant to our second example, defined in Eq.~\eqref{eq:ex_gsquared}.  In particular, at the points ${\mu = \frac{1}{2}(1\pm\sqrt{5})}$, our algorithm leads to $b_2 = 1$. Applying Eq.~\eqref{eq:recursiong} gives $g_{1}(z)=0$. Then using the above, we have that the ground state is $\ket{\psi_2}$---the cluster state. Alternatively, as shown in Eq.~\eqref{eq:ex_z2h(z)}, we have mutually inverse zeros at these values of $\mu$---this means that the ground state is given by the ground state of $f(z)=z^2$.  It is easy to generalize this observation to any $g(z)$ of degree two which has $s_0= s_2$, implying that $ b_2=1$. 

There are nevertheless models with $\lvert b_k\rvert =1$ where our approach does not work. Following on from the immediately preceding example, consider $g(z)$ with degree two and with $b_2=-1$, i.e., $s_0=-s_2$. This cannot be simplified unless $s_1=0$---then $g(z) =(z-1)(z+1)$ which is a multicritical point with the same ground state as $f(z)=-z^2$. For other values of $s_1$ we have a gapped model where our approach fails. One can argue that for these gapped models, we can define a perturbed model with $s_0=-s_2+\epsilon$, where both the ground state wave function and the corresponding Hamiltonian continuously depend on the parameter $\epsilon$ such that the limit $\epsilon \rightarrow 0$ is well-defined. However, the question remains how to take this limit.
Note that despite not having an explicit MPS in the limit, we can argue that the upper bound on $\chi$ remains valid at the limiting point. Indeed, the optimal $\chi^2$ is the number of non-zero eigenvalues of the reduced density matrix of a subsystem, and since the state is continuous, the number of non-zero eigenvalues cannot be greater at the limit point.

To see explicitly that there still exists an MPS representation with an appropriate bound on the bond dimension we can consider the positive eigenvalues of the correlation matrix \cite{Vidal2003,Peschel2003}. For a subsystem of size $N$ there will be $N$ of these eigenvalues, $\{\nu_1,\dots\nu_N\}$, where any $\nu_j=1$ is a trivial eigenvalue. The eigenvalues of the reduced density matrix can be derived from these $\{\nu_j\}$, and the number of non-zero eigenvalues of the reduced density matrix is $2^x$ where $x$ is the number of non-trivial $\nu_j$. Now, consider the model with 
\begin{align} g(z) = z^{-2}(z-z_1) \left(z-\mathcal{Z}_1\right) \label{eq:examplecorrmatrix} \end{align} where $|z_1|<1$ and $|\mathcal{Z}_1|>1$. In \cite{JV}, the eigenvalues of the correlation matrix in this model are found for any subsystem size. For a subsystem of size $N\rightarrow \infty$, there are two generically non-trivial eigenvalues given by:
\begin{align}
\nu_1^2=\nu_2^2 = {\frac{(1-z_1^2)(1-1/\mathcal{Z}_1^{2})}{(1-z_1/\mathcal{Z}_1)^2}}.
\end{align}

Now, the model in Eq.~\eqref{eq:examplecorrmatrix} has $b_2=1$ when $\mathcal{Z}_1=z_1^{-1}$. Then $\nu_1^2=\nu_2^2=1$: all correlation matrix eigenvalues are trivial. This is consistent with what we proved above, $k_0+p =0$, and thus our system has the ground state $\ket{\psi_0}$.

The model in Eq.~\eqref{eq:examplecorrmatrix} has $b_2 = - 1$ when $\mathcal{Z}_1= - z_1^{-1}$. Then 
\begin{align}
\nu_1^2=\nu_2^2 = {\frac{(1-z_1^2)^2}{(1+z_1^2)^2}}.
\end{align}
This corresponds to a bond dimension $\chi=2$ which is the same as the general case $\mathcal{Z}_1\neq\pm z_1^{-1}$ where our construction above applies. Hence, although we do not have a construction of the MPS in the case $b_2=-1$, the limiting point is an MPS with a bond dimension that is upper bounded by the path approaching it, as we expected by the general continuity argument.
\section{Outlook}
We have introduced the idea of the MPS skeleton underlying the phase diagram of one-dimensional models. To illustrate this concept, we have given a simple characterization of this skeleton for translation-invariant models in the free-fermion BDI class, as well as a construction of the MPS ground state for every model on the skeleton up to a measure-zero set of exceptions. Hamiltonians on this measure-zero set are limits of cases where we have the MPS construction; it would be interesting to see explicitly how to construct the MPS ground state in these cases. It would also be of interest to find a unitary circuit representation that applies more generally than the subset of cases we discuss above.

A natural problem is to extend our work to models in the BDI class with a larger unit cell, as well as other free-fermion classes. As discussed in Section \ref{sec:U(1)}, translation-invariant
models in class AIII are equivalent to BDI models with a two-site unit cell, so these problems are related. The characterization of the MPS skeleton for the translation-invariant AIII class is the same as in the translation-invariant BDI class, and we expect it would be relatively straightforward to adapt the construction of the MPS given in this work to this class. A more challenging generalization would be to consider Majorana chains in class D where complex hopping and pairing are allowed, or even to two-dimensional free-fermion systems, where one can analogously consider free-fermion circuits and projected entangled pair state (PEPS) skeletons. The results of Refs.~\cite{Schuch08,Kraus10} on Gaussian MPS apply for higher spatial dimensions, and make clear the following important property of MPS-solvable models: the correlation matrix in Fourier space has entries that are rational functions.
It remains to be shown that this is a sufficient condition for MPS-solvability. Perhaps this can be shown in a constructive manner; as done in the present work for the one-dimensional BDI class.
Moreover, it would be helpful to clarify the exact relationship between the construction of the ground state MPS given in this paper with the Gaussian MPS and PEPS appearing in Refs.~\cite{Schuch08,Kraus10}. This could indicate how to generalize our construction to higher dimensions. 

Finding the MPS skeleton for class of models beyond free-fermions would be very interesting. In particular, as discussed in Section~\ref{sec:results}, we found that \emph{any} MPS state in the BDI class can be obtained by starting with a fixed-point wavefunction and applying a finite number of imaginary time-evolutions generated by fixed-point Hamiltonians. An open question is whether this characterization remains true for interacting MPS.

Let us point out that the unitary circuit representation of parts of the MPS skeleton presents a powerful approach for processing quantum states on near-term quantum computers. While these states are specified by a number of parameters $b_k$ proportional to the range of the Hamiltonian, the classical processing and extraction of observables requires the contraction of matrices with bond dimension exponential in the number of parameters. The quantum circuit, however, has a unit element with its depth directly proportional to the number of parameters, leading to a possible exponential speed improvement for processing the states. This approach has been demonstrated in Ref.~\cite{Smith19} for the example discussed in Section~\ref{sec: example 1}. Whether this favourable scaling can be extended to the entire MPS skeleton is an interesting and open question; the methods in Ref.~\cite{Fishman15} could potentially be useful here. While this advantage is perhaps artificial for the exactly solvable and one-dimensional systems considered in this paper, possible generalizations may be of practical value for studying non-trivial topological phases on quantum computers. 

Finally, these exactly-solvable MPS states could serve as useful initial states which could be dressed with non-integrable perturbations. For instance, the states we construct could potentially be useful initializations for Gutzwiller-projected DMRG \cite{Jin20,Wu20,Aghaei20,Jin21,Petrica21}.
\begin{acknowledgments}
	\noindent We thank Norbert Schuch for helpful discussions and comments on the manuscript.
    F.P. and A.S. acknowledge support from the European Research Council (ERC) under the European Union's Horizon 2020 research and innovation programme (grant agreement No. 771537). 
    F.P. acknowledges the support of the Deutsche Forschungsgemeinschaft (DFG, German Research Foundation) under Germany's Excellence Strategy EXC-2111-390814868 and TRR 80. A.S. was supported by a Research Fellowship from the Royal Commission for the Exhibition of 1851. R.V. was supported by the Harvard Quantum Initiative Postdoctoral Fellowship in Science and Engineering and by the Simons Collaboration on Ultra-Quantum Matter, which is a grant from the Simons Foundation (651440, Ashvin Vishwanath).
    DMRG calculations were performed using the TeNPy Library \cite{Hauschild18}.
\end{acknowledgments}

\bibliography{arxiv.bbl}

\appendix
\section{Translation-invariant BDI Hamiltonians}\label{app:BDI}
\subsection{Analysis of  \texorpdfstring{$f(z)$}{f(z)}}
For a general translation-invariant BDI Hamiltonian given in Eq.~\eqref{eq:H}, up to an unimportant normalization, we have:
\begin{align}
f(z) &=  \sigma z^p  \prod_{\lvert z_j\rvert<1}(z-z_j)^{m_{z_j}}\prod_{\lvert \mathcal{Z}_k\rvert>1}(z-\mathcal{Z}_k)^{m_{\mathcal{Z}_k}}h(z)f_0(z)\nonumber\\
h(z) &=\sigma_h z^{-N_\zeta} \prod_{\lvert \zeta_l \rvert < 1} \left((z-\zeta_l) (z-\zeta_l^{-1})\right)^{m_{\zeta_l}} = h(1/z)\nonumber\\
f_0(z) &= \prod_{k_n \in [0,2\pi)}(z-e^{i k_n})^{m_{k_n}};\label{eq:fzcanon}
\end{align}
where the multiplicities $m$ are positive integers and $\sigma,\sigma_h \in \{\pm1\}$. Note that our Hamiltonian defines $f(z) = \sum_\alpha t_\alpha z^\alpha$, we are then putting this Laurent polynomial into a canonical form. In particular, we separate out $f_0(z)$, the zeros on the unit circle, and $h(z)$, made up of mutually inverse zeros that are not on the unit circle. The number of such mutually inverse zeros inside the unit circle, counting multiplicity, is denoted by $N_\zeta$, and it is helpful to include the factor $z^{-N_\zeta}$ so that $h(z)=h(1/z)$. By defining $h(z)$, we then have that the other zeros satisfy $z_j \neq \mathcal{Z}_k^{-1}$ for all $j,k$. Note that since the couplings $t_\alpha \in \mathbb{R}$, all zeros are real or appear in complex conjugate pairs. Due to this condition, $h(z)$ is real on the unit circle. Moreover, since $h(z)$ has no zeros on the unit circle it cannot vanish and so has a constant sign. By defining the sign $\sigma_h$ appropriately, the sign of $h(z)$ is fixed to be positive. We explain below how to do this in practice.

The Hamiltonian given in Eq.~\eqref{eq:H} can be diagonalized by modes labelled by $k \in [0,2\pi)$, the energy of each mode is given by $\lvert f(e^{i k}) \rvert$ and the mode itself is defined by the complex phase 
$f(e^{ik })/\lvert f(e^{i k}) \rvert$ \cite{Suzuki71,Keating2004,Verresen18}. In particular, the ground state depends only on this complex phase. Note that the system is gapped if and only if $f_0(z) =1$, while if $m_{k_n}\geq 1$ then at low energies we have a fermionic mode at $k_n$ with dispersion given by $\varepsilon_k \sim (k-k_n)^{m_{k_n}}$. 

\subsection{Fixing the sign of  \texorpdfstring{$h(z)$}{h(z)}}
In this section we give two characterizations of the sign of $h(z)/\sigma_h$, thus determining in a simple way the choice of $\sigma_h$ that makes $h(z)$ positive on the unit circle.

First, let
\begin{align}
h(z)/\sigma_h &= z^{-N_\zeta} \prod_{\lvert \zeta_l \rvert < 1} \left((z-\zeta_l) (z-\zeta_l^{-1})\right)^{m_{\zeta_l}} =\sum_\alpha r_\alpha z^\alpha.
\end{align}
Since $h(z)=h(1/z)$ we have $r_\alpha = r_{-\alpha}$. Then our first characterization of the sign is that \begin{align}
\textrm{sign}(h(z)/\sigma_h) = \textrm{sign}(r_0).
\end{align} This follows from the intermediate value theorem. In particular:
\begin{align}
h(e^{i k})/\sigma_h - r_0 = \sum_\alpha r_\alpha \cos(k \alpha). 
\end{align}
The right-hand side of this equation integrated over $k \in [0,2\pi)$ gives zero and hence takes both positive and negative values. By the intermediate value theorem this means the right-hand side must vanish at some value $k_0 \in [0,2\pi)$, and at that point $h(e^{i k_0})/\sigma_h= r_0$. Since the signs of each of these expressions is constant, the sign of $h(e^{i k_0})/\sigma_h$ is the same as the sign of $r_0$. 

We also have that since the sign of $h(z)/\sigma_h$ is constant on the unit circle, it must be the same as the sign of $h(1)/\sigma_h$. This is given by:
\begin{align}
h(1)/\sigma_h &= \prod_{\lvert \zeta_l \rvert < 1}(-\zeta_l^{-1})\prod_{\lvert \zeta_l \rvert < 1} \left(1-\zeta_l \right)^{2m_{\zeta_l}}.
\end{align}
Since the zeros are real or come in complex conjugate pairs, the second product is positive. Hence the sign of $h(1)/\sigma_h$ is given by the sign of the first product, which is the number of zeros of $h(z)$ outside the unit circle and on the positive real axis.

In conclusion, we can canonically choose $\sigma$ and $\sigma_h$ so that $h(z)$ has a positive sign on the unit circle, and we give two simple characterizations that fix $\sigma_h$. 
\subsection{The gapped case and correlations}
The purpose of this section is to show that the ground state is independent of $h(z)$ and, furthermore, that any translation-invariant BDI Hamiltonian outside of our class of interest cannot be represented by a finite bond dimension MPS. Suppose then that we are in the gapped case, $f_0(z)=1$. Then we have that: 
\begin{align}
\frac{f(z)}{\lvert f(z)\rvert}=\sqrt{\frac{f(z)}{f(1/z)}},
\end{align}
where the branch of the square-root is chosen so that the right-hand side has the same sign as $f(z)$ at $z=1$. We can now analytically continue this function away from the unit circle, and conclude that the ground state is determined by the function:
\begin{align}
\sigma z^{N_z+p}\!\left(\frac{\prod_{\lvert z_j\rvert<1}(1-z_j/z)^{m_{z_j}}\!\prod_{\lvert \mathcal{Z}_k\rvert>1}(1-\mathcal{Z}_k^{-1}z)^{m_{\mathcal{Z}_k}}}{\prod_{\lvert z_j\rvert<1}(1-z_jz)^{m_{z_j}}\!\prod_{\lvert \mathcal{Z}_k\rvert>1}(1-\mathcal{Z}_k^{-1}/z)^{m_{\mathcal{Z}_k}}}\right)^{\!\frac{1}{2}}\hspace{-0.25cm}, \label{eq:sqrt}
\end{align}
where $N_z$ is the total number of zeros inside the unit circle, counting multiplicity and we take the principal branch of the square-root \cite{Verresen18, Jones2019}. Notice that $h(z)$ has dropped out, so as claimed in the main text it can be ignored in deriving the ground state. Moreover, if all $m$ are even, then we are in the class analyzed in the main text with $f(z) =\sigma z^pg(z)^2h(z)$, hence, any translation-invariant BDI Hamiltonian outside of this class has at least one odd $m$.\par
\begin{figure}
\centering
\begin{minipage}{.45\textwidth}
\includegraphics{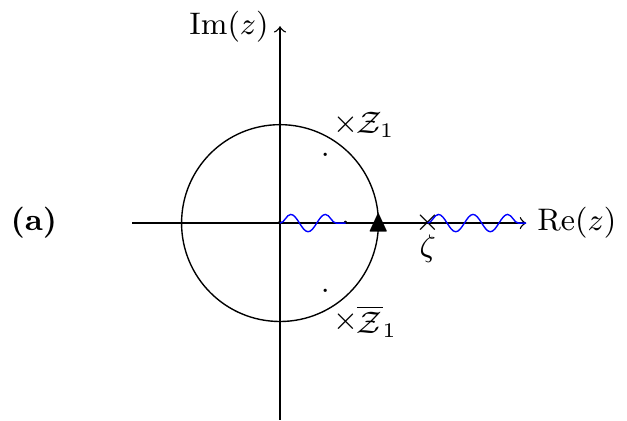}
\includegraphics{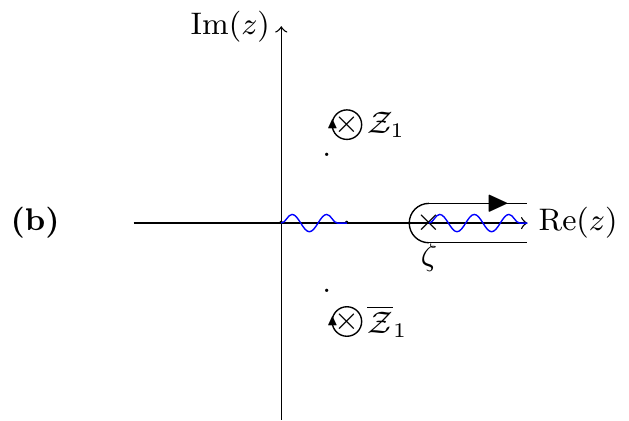}
\end{minipage}
\caption{\textbf{Contour integral for the Fourier transform.} \textbf{(a)} The contour is the unit circle, integrating over this contour defines the Fourier coefficients of \eqref{eq:sqrt}. \textbf{(b)}
The deformed contour gives the same integral and is snagged at poles and branch cuts (indicated by blue wavy lines).}
\label{fig:integral}
\end{figure}

Now, the ground state correlation function $-\langle h_{n,\alpha} \rangle$ is the $n$th Fourier coefficient of \eqref{eq:sqrt} \cite{Jones2019}. Other correlation functions can be derived from this one using Wick's theorem. The function \eqref{eq:sqrt} is analytic on the unit circle and we can compute the asymptotics of $-\langle h_{n,\alpha} \rangle$ by analytic continuation.  For $n>0$, by deforming the contour out to infinity we pick up the dominant contributions wherever the contour gets snagged: at poles and branch points. This is the Darboux principle \cite{Dingle} and is illustrated in Figure \ref{fig:integral}. The function \eqref{eq:sqrt} has poles at $\mathcal{Z}_k$ ($z_j^{-1}$) for $m_{\mathcal{Z}_k}$ ($m_{z_j}$) even, and branch points at $\mathcal{Z}_k^{-1}$ and $\mathcal{Z}_k$ ($z_j^{-1}$ and $z_j$) for $m_{\mathcal{Z}_k}$ ($m_{z_j}$) odd. Since at least one $m$ is odd, we are guaranteed to have an order two branch point outside the unit circle (behaving like $z^{m_{\mathcal{Z}_{k}}/2}$ at $z=\mathcal{Z}_k$ or $z^{-m_{z_j}/2}$ at $z=z_j^{-1}$).  Note that all zeros of $f(z)$ are either real, or come in complex conjugate pairs---for simplicity let us suppose the nearest branch point is at a real zero; for a complex conjugate pair the relevant conclusion is the same (for a related analysis see \cite{Jones2019}). By computing the contributions of the poles between the unit circle and the branch point outside the unit circle, and applying Watson's lemma for loop integrals \cite{Temme}, we have an expansion of the form:
\begin{align}
-\langle h_{n,\alpha} \rangle = &\sum_{1<\lvert z_j^{-1}\rvert<\lvert \zeta \rvert} c_{z_j} z_j^{n}+ \sum_{1<\lvert \mathcal{Z}_k\rvert<\lvert \zeta \rvert} c_{\mathcal{Z}_k} \mathcal{Z}_k^{-n}\nonumber \\&+ \frac{c_\zeta}{n^K}\zeta^{-n}(1+o(1))  \qquad n\rightarrow \infty
\end{align}
where $c_k$ are constants, $\zeta$ is the location of the branch point nearest to the unit circle and $K=1-m_{z_j}/2$ for $\zeta= z_j^{-1}$ or $K=1+m_{\mathcal{Z}_k}/2$ for $\zeta = \mathcal{Z}_k$.

In Appendix~\ref{app:multicritical} we show that critical models outside the class considered in the main text have purely algebraic terms in correlation functions. This leads to the conclusion that all translation-invariant BDI Hamiltonians outside the class considered in the main text have algebraic factors $n^{-K}$ for $K\not\in\mathbb{N}_+$ appearing in correlation functions. This is inconsistent with behaviour of correlation functions in MPS \cite{Fannes1992,Perez2007}.

\section{Quantum Ising chain} 
\subsection{Proof that the MPS path is optimal} \label{app:MPSpath}
In this section we prove that the MPS path defined by truncating the series expansion of $\sqrt{1+Jz}$ (given in Eq.~\eqref{eq:MPSpath}) is the optimal path of MPS approximations in the space of polynomials which do not have roots inside the unit disk, even allowing for a pole at zero. 

For any $\alpha,\beta \in \mathbb N$, define the following space of Laurent polynomials:
\begin{align}
\mathcal P_{\alpha,\beta} = \left\{ q(z) = \sum_{k=-\alpha}^\beta q_k z^k \; \bigg| \; q_k \in \mathbb R,~ q(z) \neq 0 \textrm{ if } |z| \leq 1 \right\}.
\end{align}
Consider $p(z) = \sum_{k=0}^\infty a_k z^k$ with $a_k \in \mathbb R$ and with radius of convergence $R>1$. We define the functional $E[~\cdot~;p]: \mathcal P_{\alpha,\beta} \to \mathbb R $ by
\begin{align}
E[q;p] = \frac{1}{4\pi i} \oint_{S^1} \frac{q(z)}{q(1/z)} p(1/z)^2 \frac{\mathrm d z}{z}.
\end{align}
Note that $-E[q;p]$ is the expectation value of the Hamiltonian determined by $f(z)=p(z)^2$ in the ground state of the Hamiltonian determined by $f(z)=q(z)^2$ (since $-\langle h_{n,\alpha}\rangle$ is the $n$th Fourier coefficient of \eqref{eq:sqrt}). In our application $p(z)^2=1+Jz$ and $q(z)$ corresponds to a Hamiltonian on the MPS skeleton. The following result proves that for the class of $q(z)$ with no zeros inside the unit circle and with degree $m$, putting $q(z)=g_m(z)$ as in the main text is optimal as claimed.
\begin{shaded}
\textbf{Proposition:} Take $p(z) = \sum_{k=0}^\infty a_k z^k$ with $a_k \in \mathbb R$ and with radius of convergence $R>1$. Consider the partial sum $q_\textrm{max}(z) := \sum_{k=0}^\beta a_k z^k$ of $p(z)$. If $q_\textrm{max}(z)$ has no roots inside the unit disk (i.e., $q_\textrm{max}(z) \in \mathcal P_{\alpha,\beta}$), then $q_\textrm{max}$ gives the global maximum of $E[q;p]$ for $q\in\mathcal P_{\alpha,\beta}$ . Moreover, $E[q_\textrm{max};p] = \frac{1}{2} \sum_{k=0}^\beta a_k^2$.

\end{shaded}

Note that if we take such a $p(z)$ with no roots inside the unit circle, then it converges uniformly and so we must have that $q_{\textrm{max}}(z)$ has no roots inside the unit circle for sufficiently large $\beta$. Moreover, when we apply this result for $p(z)^2=1+Jz$, this holds for all $\beta$.

\textbf{Proof:} For any $q(z) \in \mathcal P_{\alpha,\beta}$, let $\tilde \alpha \in \mathbb N$ be the largest integer such that $q_{-\tilde \alpha} \neq 0 $, and then let $\tilde \beta \in \mathbb Z$ be the largest integer such that $ q_{\tilde \beta}\neq0$ (note that $0 \leq \tilde \alpha \leq \alpha$ and $-\tilde\alpha \leq \tilde \beta \leq \beta$). Then we can write $q(z) = z^{-\tilde \alpha} \tilde q(z)$ with $\tilde q(z) \in \mathcal P_{0,\tilde \alpha+\tilde \beta}$. We will now show that $E$ does not depend on $a_k$ if $k > \tilde \beta - \tilde \alpha$. First note that $\partial_{a_k} p(z) = z^{k}$, thus
\begin{align}
\frac{\partial E}{\partial a_k} &= \frac{1}{4\pi i} \oint \frac{q(z)}{q(1/z)} 2 p(1/z) z^{-k} \frac{\mathrm dz}{z} \\&\propto \sum_{k'=0}^\infty a_{k'} \oint \frac{\tilde q(z)}{\tilde q(1/z)} z^{-k-k'-2\tilde \alpha-1} \mathrm dz. \label{eq:E_der}
\end{align}
Since $\tilde q(z)$ is a polynomial with no roots inside the unit disk, we know that there exists an expansion $\frac{1}{\tilde q(1/z)} = \sum_{r=0}^\infty c_r z^{-r} $ which converges on the unit circle. Hence, the largest power appearing in the integrand of Eq.~\eqref{eq:E_der} is $z^{-k-2\tilde \alpha-1+(\tilde \alpha + \tilde \beta)} = z^{\tilde \beta - \tilde \alpha - k -1}$ (note that we use that $a_0 \neq 0$, which indeed follows from the assumption that the partial sum of $p(z)$ has no root inside the unit disk). If $k > \tilde \beta - \tilde \alpha$, we thus see that there is no term proportional to $z^{-1}$ in any of the integrals in Eq.~\eqref{eq:E_der}---then, by the residue theorem, the derivative ${\partial E}/{\partial a_k}$ is always zero. 

Due to this independence, we can without loss of generality set $a_{k} = 0$ for $k > \tilde \beta- \tilde \alpha$, i.e., we truncate $p(z)$. If we choose $\tilde \alpha=0$, it is then possible to set $\tilde q(z)$ equal to this truncated $p(z)$, denoted $p_{\tilde\beta}(z)$. To see that this is indeed the optimal choice, write $\tilde q(e^{ik}) = \rho_k e^{i\phi_k}$ and $p_{\tilde\beta}(e^{ik}) = r_k e^{i\theta_k}$ (note that due to the real coefficients, we have that $\tilde q(1/z) = \rho_k e^{-i\phi_k}$ and similarly for $p_{\tilde\beta}(1/z)$), then
\begin{align}
E = \frac{1}{4\pi} \int_0^{2\pi} e^{i2\left(\phi_k-\theta_k\right)} r_k^2 \mathrm dk.
\end{align}
This is clearly maximal if and only if $\phi_k=\theta_k$, which is achieved by setting $\tilde q(z) = p_{\tilde\beta}(z)$.

Finally, the value is then given by
\begin{align}
E_\textrm{max }  &= \frac{1}{4\pi i} \oint p_{\tilde\beta}(z) p_{\tilde\beta}(1/z) \frac{\mathrm d z}{z}\nonumber\\& = \frac{1}{4\pi i} \oint \left( \sum_{k=0}^{\tilde \beta} a_k z^k \right) \left( \sum_{k=0}^{\tilde \beta} a_k z^{-k} \right) \frac{\mathrm d z}{z} = \frac{1}{2}\sum_{k=0}^{\tilde \beta} a_k^2. \label{eq:E_max}
\end{align}
Although we have a local maximum of $E$ for every allowed choice of $\tilde \beta$, Eq.~\eqref{eq:E_max} is clearly globally maximized if we choose $\tilde \beta$ as large as possible, i.e., $\tilde \beta = \beta$.
\subsection{Variational energy }\label{app:energy}
We now explain how to derive Eq.~\eqref{eq:venergy}. If we take the ground state of $f_\textrm{var}(z) = \frac{1}{z^2} ( z - z_1)^2 \left(z-\mathcal{Z}_1\right)^2$ where $|z_1|<1$ and $|\mathcal{Z}_1|>1$, then the 
energy density for the Ising Hamiltonian is given by:
\begin{align}
\mathcal{E}[z_1,\mathcal{Z}_1] = -\langle h_{n,0} \rangle -J \langle h_{n,1}\rangle.
\end{align}
As explained in Appendix \ref{app:BDI}, these expectation values are Fourier coefficients of \eqref{eq:sqrt}. These are calculated in \cite{JV} for models on the MPS skeleton. Using the result for our case, we have that $\mathcal{E}[z_1,\mathcal{Z}_1]$ is equal to:
\begin{align}
\frac{\mathcal{Z}_1 \left(z_1+\mathcal{Z}_1-J \left(z_1^2-1\right) (z_1 \mathcal{Z}_1-1)-z_1^2 \mathcal{Z}_1\right)-1}{\mathcal{Z}_1 (z_1-\mathcal{Z}_1)}.
\end{align}
We then reach Eq.~\eqref{eq:venergy} by minimizing this expression subject to $|z_1|<1$ and $|\mathcal{Z}_1|>1$ \cite{Mathematica}.

\section{Further details for the MPS representation}\label{app:MPS}

\begin{figure}[!b]
    \centering
    \includegraphics[width=\columnwidth]{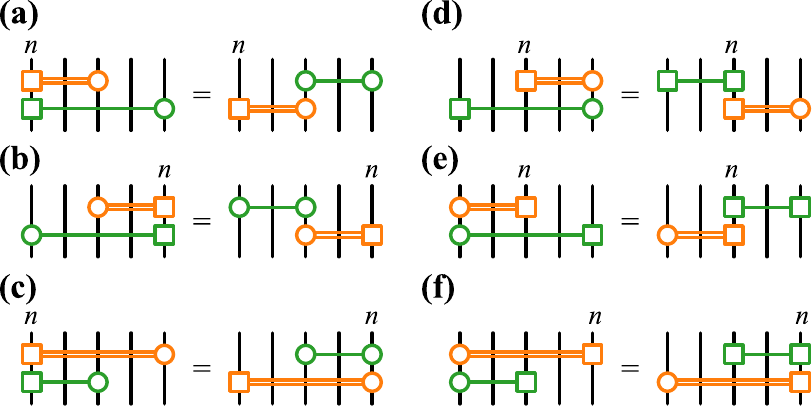}
    \caption{\textbf{Graphical representation of algebraic relations.} These relations are satisfied by the unitary gates $U_n$ (represented with double lines) and the gates $M$ (represented with single lines). These are explained explicitly in the main text. \textbf{(a-c)} are when the $\tilde{\gamma}$ operators (squares), labelled by $n$, line up. \textbf{(d-f)} correspond to when the $\gamma$ operators (circles) line up.}
    \label{fig: unitary commutation}
\end{figure}

\begin{figure*}[t!]
\includegraphics[width=\textwidth]{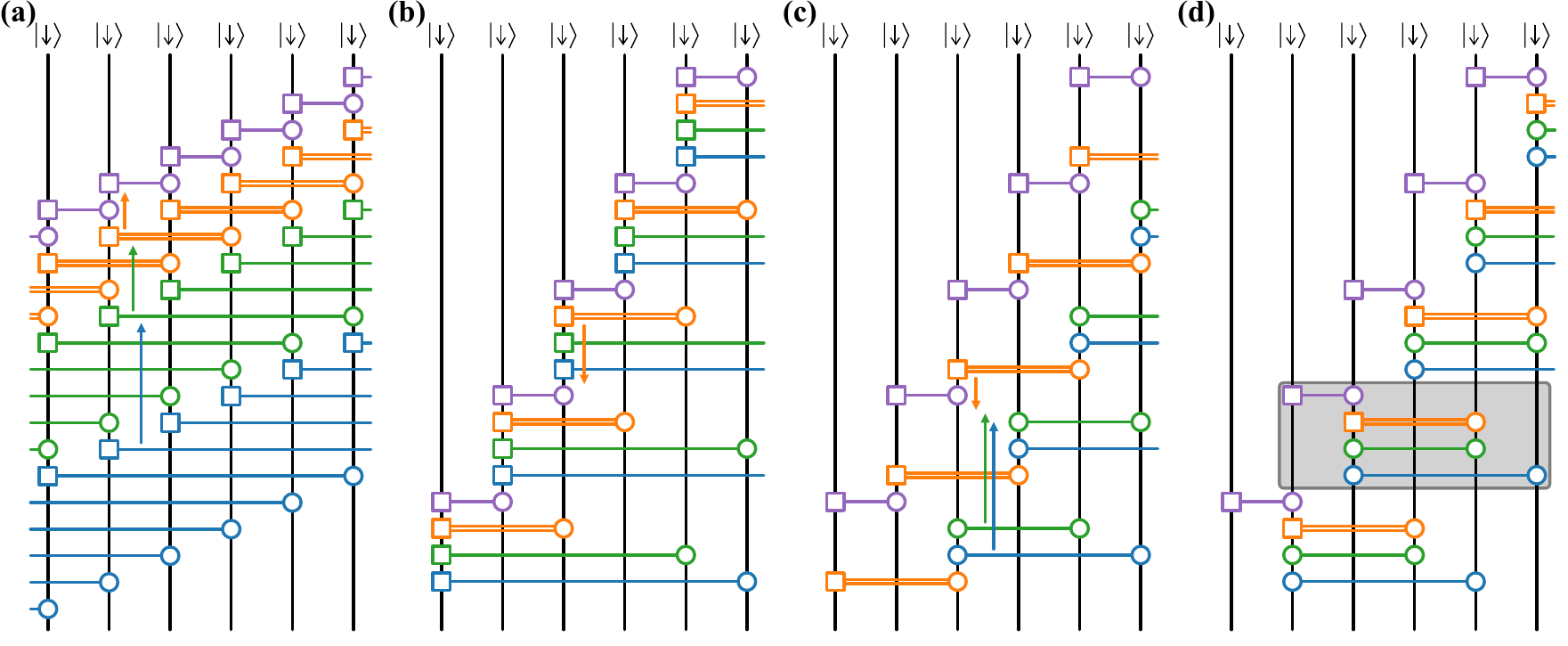}
\caption{\textbf{Explicit example for Case (i)} with $d=2$, $p=3$. }
\label{fig: explicit example 1}
\end{figure*}

\subsection{Algebraic relations for  \texorpdfstring{$U_n$}{Un}}\label{app: U commutation}

To simplify the circuits we can use some useful identities for the unitary gates $U_n$. In certain cases, bringing $U_n$ past one of the gates in $M^{(k)}$ can reduce the support of the gates and allow us to reduce the bond dimension of the equivalent MPS. 

Most generally, these relations equate to considering the unitary transformation $U^\dag_n M U^{\vphantom \dagger}_n$, where $M = 1-ic\tilde{\gamma}_a\gamma_b$ are the gates that appear in our circuit construction, and $U_n = (1-\tilde{\gamma}_n\gamma_{n+q+r})/\sqrt{2}$ are the unitary gates defined in Eq.~\eqref{eq:psip} (recall that $p=2q+r$) which we rewrite here in Majorana form for convenience. There are then two main cases: (i) when the $\tilde{\gamma}$ operators coincide with those in $U_n$ (i.e., $a=n$) as shown in Figs.~\ref{fig: unitary commutation}(a-c); (ii) when the $\gamma$ operators coincide (i.e., $b=n+q+r$) as shown in Figs.~\ref{fig: unitary commutation}(d-f). If both the $\tilde{\gamma}_a$ and $\gamma_b$ operators line up with the corresponding operator in $U_n$, then $M$ is left invariant by the unitary transformation since $U_n$ commutes with $M$.

Let us now consider these two cases explicitly. First for Case (i) when the $\tilde{\gamma}$ operators line up (and the $\gamma$ do not), then, setting $m = n+q+r$, we have
\begin{equation}
\begin{aligned}
U^\dag_n M U^{\vphantom \dagger}_n &= 1 - i \frac{c}{2} (1-\tilde{\gamma}_n \gamma_m) \tilde{\gamma}_n \gamma_b (1+\tilde{\gamma}_n \gamma_m) \\ 
& = 1 - i\frac{c}{2}(1-\tilde{\gamma}_n\gamma_m)^2 \tilde{\gamma}_n \gamma_b \\
& = 1 + ic \tilde{\gamma}_n \gamma_m \tilde{\gamma}_n \gamma_b \\
& = 1 - ic \gamma_m \gamma_b.
\end{aligned}
\end{equation}
Note, that the relative ordering of $n, m, b$ does not affect the result, so long as $m \neq b$, with two different orderings shown in Figs.~\ref{fig: unitary commutation}(a) and (c). For Case (ii) when the $\gamma$ operators line up (and the $\tilde\gamma$ do not), we have:
\begin{equation}
\begin{aligned}
    U^\dag_n M U^{\vphantom \dagger}_n &= 1 - i \frac{c}{2} (1 - \tilde{\gamma}_n \gamma_m) \tilde{\gamma}_a \gamma_m (1 + \tilde{\gamma}_n \gamma_m) \\
    &= 1 - i\frac{c}{2} (1 - \tilde{\gamma}_n \gamma_m)^2 \tilde{\gamma}_a \gamma_m \\
    & = 1 + i c \tilde{\gamma}_n \gamma_m \tilde{\gamma}_a \gamma_m \\
    & = 1 - ic \tilde{\gamma}_n\tilde{\gamma}_a.
\end{aligned}
\end{equation}
Examples are shown in Figs.~\ref{fig: unitary commutation}(d-f).

\subsection{General bond dimension formula}

In this section we run through the steps for reducing the bond dimension from the naive $\chi = 2^N$ where $N = O(d^2,p)$, to the formula 
\begin{align}
    \log_2 \chi = \lceil\textrm{range}(H)/2\rceil \label{eq:chiappendix}.
\end{align} We split this into three cases: (i) $p\geq0$; (ii) $p<0$, $d\leq|p|$; (iii) $p<0$, $d>|p|$. For each we will provide an explicit example demonstrating the steps involved in reducing the bond dimension followed by the general case.

\subsubsection{Case (i):  \texorpdfstring{$p\geq0$}{p greater than or equal to zero}}
\textbf{Example:}
Let us first consider the example with $d=2$ and $p=3$, which we show in Fig.~\ref{fig: explicit example 1} and is the same as the example in Figs.~\ref{fig: circuit to MPS 1} and~\ref{fig: circuit to MPS 2} in the main text. In this case the polynomial is of the form
\begin{equation}
    f(z) = z^3(s_0 + s_1 z + s_2 z^2)^2,
\end{equation}
which corresponds to a skeleton through the phase diagram of the Hamiltonians of the form
\begin{equation}
    H = t_3 H_3 + t_4 H_4 + t_5 H_5 + t_6 H_6 + t_7 H_7.
\end{equation}
The circuit construction for the ground state is shown in Fig.~\ref{fig: explicit example 1}(a), and is given by
\begin{equation}
    |\psi\rangle = M^{(2)} M^{(1)} U P \,|\cdots \downarrow\downarrow\downarrow \cdots \rangle.
\end{equation}
In this case, $\text{range}(H) = 7$, and so Eq.~\eqref{eq:chiappendix} states that the corresponding MPS has a bond dimension with $\log_2 \chi = \lceil 7 / 2 \rceil = 4$. However, recall that by grouping gates in the diagram to give an MPS tensor, we have a bond dimension of $\chi = 2^N$, where $N$ is the number of wires connecting unit cells. Hence,
the naive grouping of gates into a repeating unit element would give $\log_2\chi = 12$. 

The first step in simplifying the MPS representation is shown by the arrows in Fig.~\ref{fig: explicit example 1}(a), which leads to the circuit in Fig.~\ref{fig: explicit example 1}(b). This consists of commuting gates into the following form
\begin{equation}
    |\psi\rangle = \prod_n \left(M^{(2)}_n M^{(1)}_n U_n P_n \right)|\cdots \downarrow\downarrow\downarrow \cdots \rangle.
\end{equation}
This is possible using the graphical rules in Fig.~\ref{fig: graphical notation}(b), namely, that gates commute past each other so long as symbols of the same kind are not acting on the same wire. After this first step, we have already reduced the bond dimension, and by grouping with respect to the repeating element in Fig.~\ref{fig: explicit example 1}(b) we find that $\log_2 \chi = 5$ is one less than the support of $M^{(2)}_n$.

\begin{figure*}[t!]
\includegraphics[width=\textwidth]{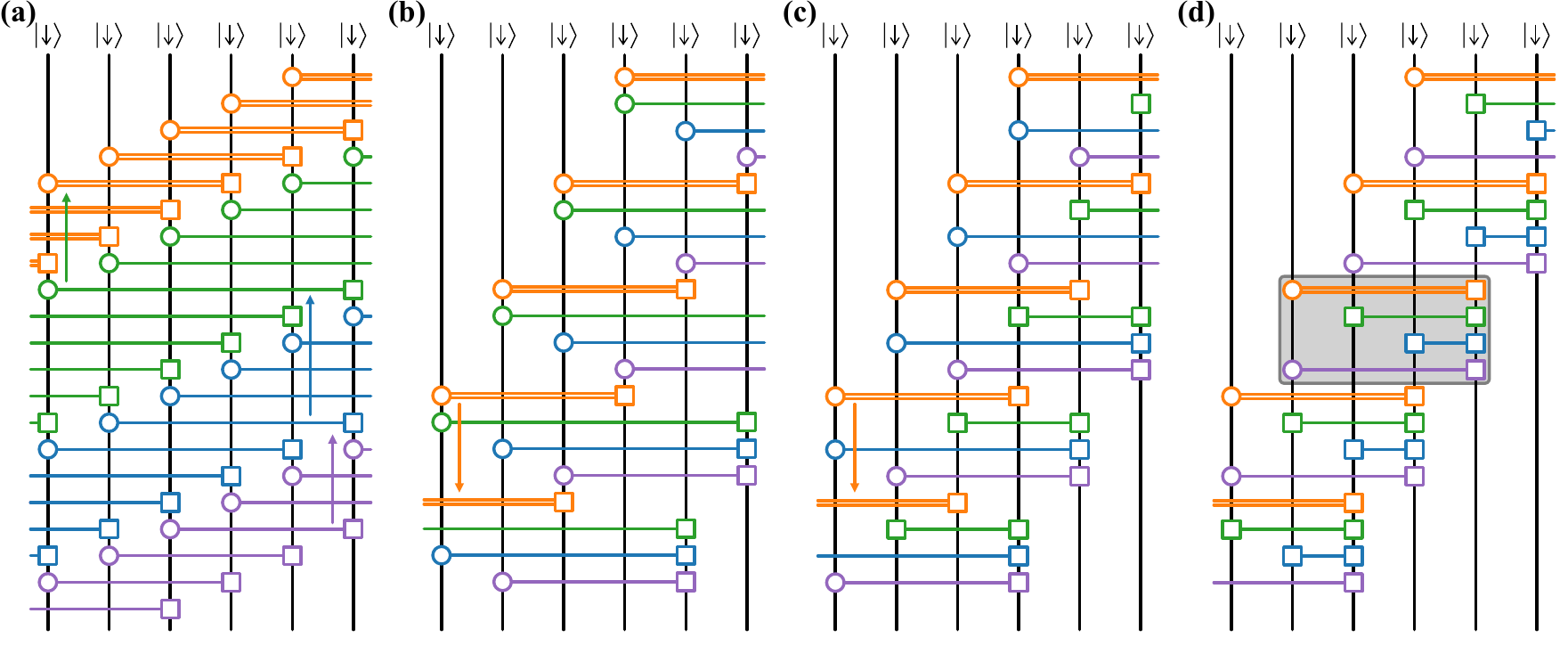}
\caption{\textbf{Explicit example for Case (ii)} with $d=3$, $p=-6$. }
\label{fig: explicit example 2}
\end{figure*}
This bond dimension can be further reduced by making use of the algebraic relations for $U_n$, described in Appendix~\ref{app: U commutation}. We use these relations to pull the unitary gate $U_n$ past $M^{(1)}_n$ and $M^{(2)}_n$ to get
\begin{equation}
    |\psi\rangle = \prod_n \left( U_n \tilde{M}^{(2)}_{n+2} \tilde{M}^{(1)}_{n+2} P_n \right) |\cdots \downarrow\downarrow\downarrow \cdots \rangle,
\end{equation}
where in this case $\tilde{M}^{(k)}_n = 1 - ia_k \gamma_n \gamma_{n+k+1}$. This is shown by the arrows in Fig.~\ref{fig: explicit example 1}(b) leading to the circuit in Fig.~\ref{fig: explicit example 1}(c). 

Finally, by using the commutation relations of the gates again, we bring the gates into the order shown in Fig.~\ref{fig: explicit example 1}(d), which corresponds to 
\begin{equation}
    |\psi \rangle = \prod_n \left( \tilde{M}^{(2)}_{n+1} \tilde{M}^{(1)}_{n+1} U_{n+1} P_n \right) |\cdots \downarrow\downarrow\downarrow \cdots \rangle.
\end{equation}
In this form we find that the bond dimension is given by $\log_2 \chi = 4$, as expected by Eq.~\eqref{eq:chiappendix}.

\textbf{General case:}
For the general case with $p\geq 0$ we can follow the same steps. First, we are able to commute gates into the form  \begin{equation}
    |\psi\rangle = \prod_n \left(M^{(d)}_n\dots M^{(2)}_n M^{(1)}_n U_n \left(P_n\right) \right)|\cdots \downarrow\downarrow\downarrow \cdots \rangle. \label{eq:initial}
\end{equation}
We place $P_n$ in brackets to indicate that it is only there in the case that $p$ is odd. Then, we use algebraic relations to pull $U_n$ past the layers of $M^{(k)}_n$, reducing the support of these gates. Finally we simply commute gates so that we can group gates by wire, where the left-most operator of all gates \emph{except} $P_n$ appears on a single wire, preceded by the right-most operator of $P_n$ (if it is there).

Let us now argue how the bond dimension changes when we do this. Firstly, when $p\geq 0$ we have $\text{range}(H) = 2d + p$. Consider the initial grouping of gates Eq.~\eqref{eq:initial}, with general $d$. The support of $M^{(d)}_n$ is $d + p + 1$, and we see that the bond dimension of  Eq.~\eqref{eq:initial} satisfies $\log_2\chi=d+p$. The support of $U_n$ is $\lceil p/2 \rceil$, and so by using the algebraic relations for $U_n$ as described, we end up with $\tilde{M}^{(d)}$ with support $d+p+1-\lceil p/2 \rceil$. Finally, if $p$ is odd then we have to include the projectors $P_n$ which increase the support of the unit circuit element by $1$. Therefore the total support of the reduced circuit is $d + p + 1 - \lfloor p/2 \rfloor$. The bond dimension is one less than this since one of the wires is physical (or projected on~$\ket{\downarrow}$). In conclusion,  we have an MPS construction with 
\begin{equation}
    \log_2 \chi = d + \lceil p/2 \rceil = \lceil \text{range}(H)/2 \rceil.
\end{equation}

\subsubsection{Case (ii):  \texorpdfstring{$p<0$, $d\leq|p|$}{p less than zero and d less than or equal to the absolute value of p}}

\textbf{Example:}
Let us now consider $d=3$ and $p=-6$, that is, a polynomial of the form
\begin{equation}
    f(z) = z^{-6}(s_0 + s_1 z + s_2 z^2 + s_3 z^3)^2,
\end{equation}
which corresponds to a skeleton through the phase diagram of the Hamiltonians of the form
\begin{equation}
    H = \sum_{k=-6}^{0} t_k H_k.
\end{equation}
The circuit construction for the ground state is shown in Fig.~\ref{fig: explicit example 2}(a). In this case, $\text{range}(H) = 6$, and so Eq.~\eqref{eq:chiappendix} states that the corresponding MPS has a bond dimension with $\log_2 \chi = \lceil 6 / 2 \rceil = 3$. However, the naive grouping of gates into a repeating unit element would give $\log_2\chi = 15$.

Similarly to the previous example, the first step is to commute the gates past each other to get the grouping in Fig.~\ref{fig: explicit example 2}(b), reducing to $\log_2\chi = 5$. Then we use the algebraic relations for the unitary gate $U_n$ to bring it past the corresponding $M^{(1)}_{n'}$, as shown going between Fig.~\ref{fig: explicit example 2}(b) and (c). (To be precise, $n'$ is the site such that the circles or $\gamma$ operators of $U_n$ and $M^{(1)}_{n'}$ agree.) Bringing $U_n$ past $M^{(1)}_{n'}$ (shown in green) corresponds to the case in Fig.~\ref{fig: unitary commutation}(e). The gate $U_n$ then commutes past $M^{(2)}_{n'}$ (shown in blue) and $M^{(3)}_{n'}$ (purple). This gives us $\log_2\chi = 4$. To go from Fig.~\ref{fig: explicit example 2}(c) to (d) we again use the relations for $U_n$, but now to bring it past $M^{(2)}_{n'-1}$. After these relations and commutation of gates the repeating unit element shown in Fig.~\ref{fig: explicit example 2}(d) corresponds to a bond dimension $\log_2 \chi = 3$, agreeing with Eq.~\eqref{eq:chiappendix}.

\textbf{General case:}
In the general case $p< 0$ and $d\leq |p|$, we proceed as follows. First, similarly to the previous case, we commute gates past each other such that we group the gates into unit cells:
\begin{equation}
    |\psi\rangle = \prod_n \left(M^{(d)}_n\dots M^{(2)}_n M^{(1)}_n U_{n-x} \left(P_{n-\lvert p\rvert}\right) \right)|\cdots \downarrow\downarrow\downarrow \cdots \rangle, \label{eq:initialcase2}
\end{equation}
where\footnote{Since $p<0$, if $p$ is even then we apply $W_{-\lfloor \lvert p \rvert/2 \rfloor}$, while if $p$ is odd we apply $W_{-\lfloor \lvert p \rvert/2 \rfloor}P$.} $x=\lceil |p| / 2 \rceil -1$.
Next, we repeatedly use the algebraic relations for $U_n$, as well as the commutation relations, to reduce the support of the gates appearing in $M^{(k)}$. In particular, we bring each $U_{n-x}$ past $M^{(d)}_m\dots M^{(2)}_m M^{(1)}_m$ for $n \geq m >n-x$, then Eq.~\eqref{eq:initialcase2} becomes:
\begin{align}
     |\psi\rangle =\prod_n \Big(M^{(d)}_n \dots M^{(\lceil |p|/2\rceil)}_n \tilde{M}^{(\lceil |p|/2 \rceil-1)}_n\dots\nonumber\\\dots \tilde{M}^{(2)}_n \tilde{M}^{(1)}_n U_{n} \left(P_{n-\lfloor |p|/2\rfloor-1}\right) \Big)|\cdots \downarrow\downarrow\downarrow \cdots \rangle. \label{eq:case2}
\end{align}

Let use now consider how these steps change the bond dimension. In this case, we have $\text{range}(H) = |p|$. The gates $M^{(k)}_n$ have support $|k + p| + 1 = 1 - p - k$. However, using the algebraic relations for $U_n$, we find that if this support is greater than that of $U_n$ (which is $\lfloor |p|/2 \rfloor + 1$) then $\tilde{M}^{(k)}$ has support less than $U_n$. Now, the support of $M^{(k)}$ is greater than that of $U_n$ for $k< \lceil |p|/2\rceil$. Note that we only include $P_{n-\lfloor |p|/2\rfloor-1}$ if $p$ is odd. Hence,
by considering Eq.~\eqref{eq:case2} we see that the support of the reduced circuit is $\lceil |p| / 2 \rceil + 1$ and so 
\begin{equation}
    \log_2 \chi = \lceil |p| /2 \rceil = \lceil \text{range}(H)/2 \rceil.
\end{equation}

\begin{figure*}[t!]
\includegraphics[width=0.75\textwidth]{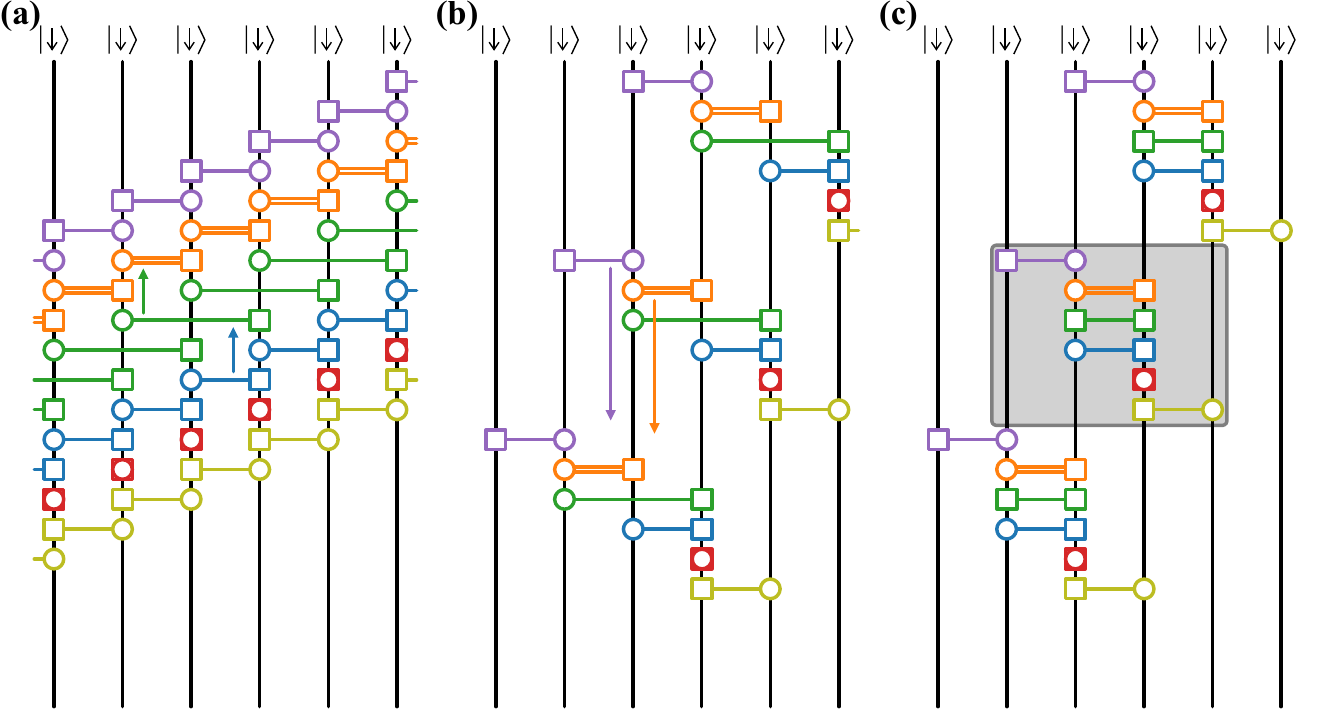}
\caption{\textbf{Explicit example for Case (iii)} with $d=4$, $p=-3$. }
\label{fig: explicit example 3}
\end{figure*}

\subsubsection{Case (iii):  \texorpdfstring{$p<0$, $d>|p|$}{p less than zero and d greater than the absolute value of p}}
\textbf{Example:}
For the final illustrative example, we consider $d=4$, and $p=-3$. This corresponds to a polynomial of the form
\begin{equation}
f(z) = z^{-3}(s_0 + s_1 z + s_2 z_2 + s_3 z^3 + s_4 z^4)^2
\end{equation}
and to the Hamiltonian
\begin{equation}
    H = \sum_{k=-3}^{5} t_k H_k.
\end{equation}
In this case, $\text{range}(H) = 5$ and so Eq.~\eqref{eq:chiappendix} states that the MPS has bond dimension $\log_2\chi = \lceil 5/2 \rceil = 3$. The circuit construction for the ground state is shown in Fig.~\ref{fig: explicit example 3}(a), which has a naive grouping of gates with unit element $\log_2\chi = 6$.

The first step in reducing this bond dimension is to use the commutation relations to bring the gates together in the form shown in Fig.~\ref{fig: explicit example 3}(b). This has a reduced bond dimension of $\log_2\chi = 4$. Next we bring the unitary $U_n$ past the gate $M^{(1)}$ (shown in green) to go from Fig.~\ref{fig: explicit example 3}(b) to (c). $U_n$ then commutes past the rest of the $M^{(k)}$ gates, as does the initial projector $P$ (shown in purple). This brings the circuit into the final form shown in Fig.~\ref{fig: explicit example 3}, which has a repeating unit element with support over 4 sites. This means that the resulting bond dimension is $\log_2\chi = 3$, in agreement with Eq.~\eqref{eq:chiappendix}.

\textbf{General case:}
The steps for the general case proceed similarly to when $p<0$ and $d\leq |p|$. Again, we begin by grouping into a unit cell, each centred around a ``spine'' of $\tilde{\gamma}$ operators (squares). This leads to exactly the formula given in Eq.~\eqref{eq:initialcase2}. Moreover, using the commutation relations as in the previous case, we reach Eq.~\eqref{eq:case2}. The difference with this case is that, for example, $M_n^{(d)}$ extends to the right of the spine. Thus the support of the reduced circuit can be found by considering $M_n^{(d)}$---which is supported between sites $n$ and $n+d+p$---and either $U_{n}$ for $p$ even or $U_{n} P_{n-\lfloor |p|/2\rfloor-1}$ for $p$ odd---which in both cases is supported between sites $n$ and $n-\lceil |p|/2\rceil$. 
Therefore we have that 
\begin{equation}
\log_2\chi = d - \lfloor |p| / 2 \rfloor = \lceil \text{range}(H) / 2 \rceil.
\end{equation}
\section{Calculations for special cases}

\subsection{\texorpdfstring{$U(1)$}{U(1)} symmetric chains } \label{app:U(1)}
Let us consider the fermionic hopping chain with sublattice symmetry defined in Eq.~\eqref{eq:H U(1)}, with complex $\tau_\alpha$. We can define Majorana operators as follows:
\begin{align}
2c_{A,n} &= -\gamma_{2n}+i \gamma_{2n-1}\\
2c_{B,n} &= \tilde\gamma_{2n-1}+i \tilde\gamma_{2n}.
\end{align} 
Then the Hamiltonian given in Eq.~\eqref{eq:H U(1)} is transformed to $H_R$ + $H_I$, where these are defined as:
\begin{align}
H_R&= \frac{i}{2}\sum_{\alpha, n} \Re(\tau_\alpha)\tilde{\gamma}_{n}\gamma_{n+2\alpha} \label{eq: U(1) majorana}
\\
H_I&=\frac{i}{2}\sum_{\alpha, n} \Im(\tau_\alpha)\left(-\tilde{\gamma}_{2n-1}\gamma_{2n+2\alpha}+\tilde{\gamma}_{2n}\gamma_{2n+2\alpha-1} \right).
\end{align}
Hence, if $\tau_\alpha \in \mathbb{R}$ then this is equivalent to a translation-invariant BDI model as studied in the main text. The construction of the MPS given in the main text follows from considering $f(z) = \sum_{\alpha} \tau_\alpha z^{2\alpha}$ as a Majorana chain, applying Result~\ref{result2}, and then transforming back into a complex fermionic description.

\subsection{The gapless case: multicritical points}\label{app:multicritical}
In this section we analyze gapless models that fit our ansatz, i.e., $f(z) = \sigma z^p g(z)^2 h(z)$. This means that all zeros on the unit circle have even multiplicity, and so we can simplify the contribution from $f_0(z)$. In particular, we will show that the ground states of all of these gapless models are ground states of gapped models that themselves can be analyzed with our methods. This supports Section \ref{sec:MCP}.\par
In our ansatz, $f_0(z)$ must take the following form:
\begin{align}
(z-1)^{m_0} (z+1)^{m_\pi}\prod_{k_n \in (0,\pi)}\left((z-e^{i k_n})(z-e^{-i k_n})\right)^{m_{k_n}}; \label{eq:critical}
\end{align}
where the multiplicities $m_{k_n}$ are even non-negative integers. Now, to understand the ground state we need to calculate
\begin{align}
\frac{f(z)}{\lvert f(z)\rvert}=\frac{f_0(z)}{\lvert f_0(z)\rvert} \times \eqref{eq:sqrt}.
\end{align}
Using the following substitutions, valid for even integers $m_{k_n}$:
\begin{align}
\frac{(z-1)^{m_0}}{\lvert(z-1)^{m_0}\rvert} &= (-1)^{m_0/2}z^{m_0/2}  \\
\frac{(z+1)^{m_\pi}}{\lvert(z+1)^{m_\pi}\rvert} &= z^{m_\pi/2}  \\
\frac{\left((z-e^{i k_n})(z-e^{-i k_n})\right)^{m_{k_n}}}{\lvert\left((z-e^{i k_n})(z-e^{-i k_n})\right)^{m_{k_n}}\rvert}&=z^{m_{k_n}},
\end{align}
we can make the following conclusion. The model defined by
\begin{align}
f_{\textrm{gl}}(z) &= \sigma z^p  \!\prod_{\lvert z_j\rvert<1}\!(z-z_j)^{m_{z_j}}\!\prod_{\lvert \mathcal{Z}_k\rvert>1}\!(z-\mathcal{Z}_k)^{m_{\mathcal{Z}_k}}h(z)f_0(z) 
\end{align}
where all zeros apart from those in $h(z)$ have even degeneracy, has the same ground state as the model defined by
\begin{align}
f_{\textrm{g}}(z) &= \sigma (-1)^{m_0/2} z^{p+N_c/2} \nonumber\\&\times \prod_{\lvert z_j\rvert<1}(z-z_j)^{m_{z_j}}\prod_{\lvert \mathcal{Z}_k\rvert>1}(z-\mathcal{Z}_k)^{m_{\mathcal{Z}_k}},
\end{align}
where $N_c/2 \in \mathbb{Z}$ is half the number of zeros of $f(z)$ on the unit circle.  Note that $f_{\textrm{g}}(z)$ is of an appropriate form to apply the methods of the main text to find the ground state MPS.

Let us now consider Eq.~\eqref{eq:critical}, but now with some multiplicity odd. This means we must have jump discontinuities of $f(z)/\lvert f(z)\rvert$ on the unit circle. Then due to this jump discontinuity,  $-\langle h_{n,\alpha} \rangle$ behaves like $1/n$. This completes the argument in Appendix~\ref{app:BDI} that showed any model outside our class does not have an exact MPS ground state. 

\subsection{Consequences of  \texorpdfstring{$b_{k}=\pm 1$ for $g_{k}(z)$}{bk=±1 for gk(z)}}\label{app:bk=1}

Here we prove some of the results asserted in Section~\ref{sec:specialc}. In particular, the case where our method works for $b_{k_0}=\pm 1$ is a case where the state before applying the projector is a ground state of $f(z) = \pm z^{k_0}$. In the following we argue that this is equivalent to $g_{k_0-1}(z)=0$. Note that here we assume that $p=0$, we can account for general $p$ as in the main text by shifting $h_{n,\alpha}\rightarrow h_{n,\alpha+p}$.

First, suppose that the condition $g_{k_0-1}(z)=0$  holds, then rearranging Eq.~\eqref{eq:recursiong} gives 
\begin{align} 
g_{k_0}(z)/g_{k_0}(1/z) = b_{k_0} z^{k_0} 
\label{eq:rearrange}
\end{align}
which has the same ground state as $f(z) = \pm z^{k_0}$. (We show below that this sign is in fact given by $b_{k_0}$.)

Suppose that $g_{k_0}(z)$ is such that ${f(z)= g_{k_0}(z)^2}$ has the same ground state as $f(z) =\pm z^{k_0}$ according to the various simplifications discussed in the main text. Then, the most general form $g_{k_0}(z)$ can have is
\begin{align}
    g_{k_0}(z) = z^q h(z) u(z); \label{eq:defgk0}
\end{align}
here $h(z)$ is a function without zeros on the unit circle that satisfies $h(1/z) = h(z)$ and has a positive constant term, and $u(z)$ accounts for possible zeros on the unit circle. 
We can write $u(z)$ as the product:
\begin{align}
    u(z)\!=\!(z-1)^{\frac{m_0}{2}}\!(z+1)^{\frac{m_\pi}{2}}\hspace{-0.3cm}\prod_{k_n\in(0,\pi)}\hspace{-0.35cm}\left((z-e^{ik_n})(z-e^{-ik_n})\right)^{\frac{m_{k_n}}{2}}\hspace{-0.3cm},
\end{align}
with multiplicities $m_{k_j}/2\in\mathbb{Z}$. The number of zeros of $u(z)$ on the unit circle, counting their multiplicity, is $N_c/2\in\mathbb{Z}$. This notation is consistent with $f_0(z)=u(z)^2$ in Eq.~\eqref{eq:fzcanon}. Note that \begin{align}
    u(1/z)=(-1)^{m_0/2}z^{-N_c/2}u(z).
\end{align}
Then, according to Section~\ref{sec:MCP} and Appendix~\ref{app:BDI}, to find the ground state we can ignore $h(z)$ and can substitute $u(z)^2$ by $(-1)^{m_0/2}z^{N_c/2}$. This transforms ${f(z) = g_{k_0}(z)^2}$ to $f(z) = (-1)^{m_0/2}z^{k'}$ with ${k' = 2q+N_c/2}$. From Eq.~\eqref{eq:defgk0}, we have that $k' =k_0$.
Finally, note that \begin{align}
     g_{k_0}(z) = (-1)^{m_0/2} z^{k_0} g_{k_0}(1/z),
    \label{eq:gk0inv}
\end{align}
   from which we deduce that $b_{k_0}=s_{k_0}/s_0=(-1)^{m_0/2}$. Using this, inserting $g_{k_0}(z)$ into Eq.~\eqref{eq:recursiong} gives
\begin{align}
    g_{k_0-1}(z) =  g_{k_0}(z) \left(1 - (-1)^{\frac{m_0}{2}} b_{k_0}\right)
    =0.
    \label{eq:gk01}
\end{align}
Thus, we have proved our claim. Moreover, we have shown that given $b_{k_0}=s_{k_0}/s_0=\pm 1$ then we have the ground state of $f(z) = b_{k_0} z^{k_0}$. Given the form of the projector, i.e., $P^{(k_0)}=\prod_{n} \left(1+b_{k_0}h_{n,k_0}\right)$,
this also explains why we only observe the case where the projector annihilates the state. 
Note if we start from a negative global sign of $f(z)$, i.e., $f(z)=-g_{k_0}(z)^2$, then the same applies because the sign change also appears in the projector $P^{(k_0)}=\prod_{n} \left(1-b_{k_0}h_{n,k_0}\right)$.
\end{document}